 \documentclass[article,5p,times,twocolumn]{elsarticle}

\usepackage{amssymb}

\graphicspath {{figures/}}

\usepackage{amsmath,amssymb,amsfonts}
\usepackage{algorithmic}
\usepackage{graphicx}
\usepackage{bm}
\usepackage{subfig}
\usepackage{caption}
\usepackage{balance}
\usepackage{mathtools}
\usepackage{booktabs}
\usepackage{placeins}
\usepackage{float}
\usepackage{changepage}
\usepackage{tabularx}
\usepackage{multirow}
\usepackage{makecell} 
\usepackage{bigstrut}
\usepackage{booktabs}
\usepackage{float}
\usepackage{orcidlink}
\usepackage{adjustbox}
\usepackage{afterpage}
\usepackage{balance}

\usepackage{caption}

\usepackage{titlesec}
\titlespacing*{\section}{0pt}{0.3\baselineskip}{0.3\baselineskip}

\newcommand{\orcidNG}{0000-0002-8099-0627} 
\newcommand{\orcidJGM}{0000-0002-7422-5320} 
\newcommand{\orcidEV}{0000-0002-1627-6883} 
\newcommand{\orcidSLC}{0000-0003-3605-7351} 
\newcommand{\orcidMVO}{0000-0002-7680-3980} 
\newcommand{\orcidELP}{0000-0002-2743-1033} 
\newcommand{\orcidLCM}{0009-0009-3915-1707} 

\DeclareTextComposite{\.}{T1}{i}{`\i}
\DeclareTextComposite{\.}{T1}{\i}{`\i}

\journal{?????}

\begin{document}

\begin{frontmatter}



\title{EigenHearts: Cardiac Diseases Classification Using EigenFaces Approach}


\author[inst1,inst2]{Nourelhouda Groun\orcidlink{\orcidNG}}

\affiliation[inst1]{organization={ETSI Aeronáutica y del Espacio - Universidad Politécnica de Madrid},
            addressline={ Pl. del Cardenal Cisneros, 3}, 
            postcode={28040}, 
            state={Madrid},
            country={Spain}}

\affiliation[inst2]{organization={ETSI Telecomunicación - Universidad Politécnica de Madrid},
            addressline={Av. Complutense, 30}, 
            postcode={28040}, 
            state={Madrid},
            country={Spain}}

\author[inst3,inst4]{Mar\'{i}a Villalba-Orero\orcidlink{\orcidMVO}}

\affiliation[inst3]{organization={Departamento de Medicina y Cirugía Animal, Facultad de Veterinaria - Universidad Complutense de Madrid},
            addressline={Av. Puerta de Hierro}, 
            postcode={28040}, 
            state={Madrid},
            country={Spain}}
            
\affiliation[inst4]{organization={Centro Nacional de Investigaciones Cardiovasculares (CNIC) },
            addressline={C. de Melchor Fernández Almagro, 3}, 
            postcode={28029}, 
            state={Madrid},
            country={Spain}}

\author[inst3]{Luc\'{i}a Casado-Mart\'{i}n\orcidlink{\orcidLCM}}

\author[inst4]{Enrique Lara-Pezzi \orcidlink{\orcidELP}}

\author[inst1,inst5]{Eusebio Valero \orcidlink{\orcidEV}}

\author[inst1,inst5]{Soledad Le Clainche \orcidlink{\orcidSLC}}

\author[inst1,inst5]{Jes\'us Garicano-Mena \orcidlink{\orcidJGM}}

\affiliation[inst5]{organization={Center for Computational Simulation (CCS)},
            addressline={Boadilla del Monte}, 
            postcode={28660}, 
            state={Madrid},
            country={Spain}}

\begin{abstract}

In the realm of cardiovascular medicine, medical imaging plays a crucial role in accurately classifying cardiac diseases and making precise diagnoses. However, the field faces significant challenges when integrating data science techniques, as a significant volume of images is required for these techniques on the one hand, where ethical constraints, high costs, and variability in imaging protocols can limit the acquisition of sufficient amounts of medical data on the other hand. As a consequence, it is necessary to investigate different avenues to overcome this challenge. In this contribution, we offer an innovative tool to conquer this limitation. In particular, we delve into the application of a well recognized method known as the EigenFaces approach to classify cardiac diseases. This approach was originally motivated for efficiently representing pictures of faces using principal component analysis, which provides a set of eigenvectors (aka eigenfaces), explaining the variation between face images. As this approach proven to be efficient for face recognition, it motivated us to explore its efficiency on more complicated data bases. In particular, we integrate this approach,  with convolutional neural networks (CNNs) to classify echocardiography images taken from mice in five distinct cardiac conditions (healthy, diabetic cardiomyopathy, myocardial infarction, obesity and TAC hypertension). Performing a pre-processing step inspired from the eigenfaces approach on the echocardiography datasets,  yields sets of pod modes, which we will call eigenhearts. To demonstrate the proposed approach, we compare two testcases: (i) supplying the CNN with the original images directly, (ii) supplying the CNN with images projected into the obtained pod modes. The results show a  substantial and noteworthy enhancement when employing SVD for pre-processing, with classification accuracy increasing by approximately 50\%.     
\end{abstract}




\end{frontmatter}


\section{Introduction}
The contributing in the diagnosis of different disease through classification is one of
the most useful practices of data science in medicine. Nonetheless, this approach commonly encounters the hurdle of obtaining adequately large, well-curated, and representative datasets. Acquiring ample data presents one of the primary hurdles for data scientists. Consequently, they have begun exploring novel methods to tackle this issue, aiming to leverage existing data and attain the most precise classifications possible.\\

In recent years, researchers have developed a variety of techniques to address these challenges, as we can mention: (i) creating simulated or synthetic data. This last involves the use of methods such as probabilistic models \cite{khader2023denoising, shibata2024practical},  classification-based imputation models \cite{goncalves2020generation, ganesan2022data}. (ii) Few-shot learning \cite{cai2020few, singh2021metamed, dai2023pfemed}, which is mainly used to teach a model to generalize for new tasks or problems with only a few labeled examples per class. This  technique directly targets small data problems, speeds model adaptation and enhance generalisability. (iii) Ensemble learning \cite{muller2022analysis, sreelakshmi2023m, suk2017deep}, which focuses on combining many individual models that learned differently from each other for prediction and it provides  robustness for data perturbation.\\

Along side the previously mentioned tools, data augmentation is one of the most clever and commonly used approaches to maximize the size of data. Data augmentation primarily entails making alterations to the original databases to produce additional samples. This process is designed to assist the model in avoiding overfitting to the specific features of the original dataset, thereby enhancing the model's ability to better generalize new testing data. Data augmentation methods can be simple and basic techniques, such as: (i)  geometric transformations \cite{jamaludin2017spinenet}, which involves rotations, translation, rescaling and random flipping. (ii) Filtering techniques \cite{sajjad2019multi}, eg. filters such as Gaussian blur, sharpening, edge detection, and emboss. Other basic augmentation tools include cropping \cite{drozdzal2018learning, afzal2019data, hua2020lymph}, occlusion \cite{kompanek2019volumetrie} and intensity operations \cite{chen2019computer, sultan2019multi, winkels2019pulmonary}. However, if standard augmentation methods do not yield any enhancements, advanced augmentation techniques may be worth considering. Approaches such as elastic deformation \cite{lorenzo2019segmenting, karani2021test}, interpolation \cite{payer2019integrating,kim2019evcmr}, deformable image registration\cite{krivov2018mri, tustison2019convolutional} and statistical shape models \cite{corral2019smod, bhalodia2018deep}, are highly used and demonstrate a noticeable enhancement in classification performance. Another well known  data augmentation tool, which is based on deep learning is generative adversarial networks (GANs) \cite{goodfellow2014generative}. This tool is considered the most used methods for data augmentation for medical image classification problems. GANs have been used to enhance heart disease prediction \cite{gayathri2024enhancing}, cardiovascular abnormality classification \cite{madani2018chest}, as well as improve the classification of brain conditions \cite{ma2020differential}, for the diagnosis of Parkinson’s disease \cite{kaur2021diagnosis} and diagnose and classify femoral neck fractures \cite{mutasa2020advanced}. The various positive results obtained from augmenting the database using GANs highlighted its potential to be used for radiography-based deep learning projects. \\

Another common, more frequently used  strategy to address the constraint of limited data is transfer learning \cite{torrey2010transfer}. This approach has made a major contribution to medical image analysis as it overcomes the data scarcity problem. One frequently employed method in transfer learning involves initially training a neural network on a source domain, such as ImageNet, a vast image database comprising over fourteen million annotated images spanning more than 20,000 categories. Subsequently, the network is fine-tuned using data from the target domain. Several researches have explored the advantages of this approach to address the limitations of medical data. Lopes \textit{et al.} \cite{lopes2021improving} employed transfer learning to improve the detection of rare genetic heart disease. Their approach shows improvement in the detection of the disease in both balanced and imbalanced datasets. Pathak  \textit{et al.} \cite{pathak2022ensembled} have also investigated the use of transfer learning for prediction of Coronary Artery Disease Detection using limited amounts of medical data. Maqsood \textit{et al.} \cite{maqsood2019transfer} pretrained AlexNet \cite{krizhevsky2012imagenet} over ImageNet to detect Alzheimer’s disease. In particular, the convolutional layers of AlexNet are fixed, and the last three fully connected layers are replaced by the new ones and the modified AlexNet is finetuned by training on the Alzheimer’s dataset. The results show the highest accuracy for the multi-class classification problem. 
Tang \textit{et al.} \cite{tang2019combining} proposed a novel approach combining active learning (AL) and transfer learning (TL) for medical data classification. To verify the effectiveness of their algorithm, a comparative studies on ten datasets was carried out, and the experimental results confirm the importance and effectiveness of the ATL combination. \\

Besides machine learning based methods, recently several researchers have investigated the use of matrix and tensor factorization methods to address the limitations of medical data. In particular, data-driven approaches have been making a clear progress in the medical imaging field, as researchers have used these techniques not only to create synthetic electronic medical records \cite{buczak2010data}, simulation and medical image generation \cite{frangi2018simulation} and  simulation and visualization of the heart \cite{poyart2016parallel}, but also joined with Vision Transformer (ViT) to enhance the classification of cardiac pathologies \cite{BELLNAVAS2024125849}.\\
During our research, and in our efforts to classify various cardiac conditions, we have decided to inspect a famous matrix factorization tool, which is the  \textit{eigenfaces} approach \cite{sirovich1987low, turk1991eigenfaces, brunton2022data}. This technique is one of the most striking demonstrations of the singular value decomposition (SVD) \cite{golub1971singular}, where the principal component analysis (PCA) \cite{hotelling1933analysis,jolliffe2005principal,sutherland2009combustion},  is applied on substantial collection of facial images to identify the most significant correlations among them. This decomposition yields a series of \textit{eigenfaces}, which establish a novel coordinate system that can be used to represent images in these coordinates by taking the dot product with each of the principal
components. The eigenfaces approach has proven to be useful for facial recognition and
classification, as images of the same person tend to easily cluster
in the eigenface space. As a consequence, we have decided to explore this approach for the classification of medical images. \\

In more details, we aim on classifying five different cardiac conditions: healthy, diabetic cardiomyopathy, myocardial infarction, obesity and TAC hypertension using echocardiography images. However, instead of using the echocardiography images directly, we perform a pre-processing step inspired from the \textit{eigenfaces} approach. In particular, SVD is applied first to our mean-subtracted  database, which is a collection of echocardiography images of the different cardiac conditions. The  decomposition yields a set of pod modes, which we will call \textit{eigenhearts}. These modes will be used to approximately represent the original echocardiography images, by projecting them onto a new coordinate system defined by the eigenhearts. The new, projected images will be used to train a convolutional neural network in order to classify five distinct cardiac conditions. The results obtained, which will be detailed further in the paper, show a significant improvement in the classification accuracy, demonstrating the efficiency of the proposed approach.\\

The remaining of the article is organized as follows: section \ref{Methd} will introduce the methods used. The materials and data pre-processing are explained in section \ref{material}. The different results obtained are presented in section \ref{Results} and section \ref{concl} will hold the conclusions. 

\section{Methods} \label{Methd}

\subsection{Singular value decomposition (SVD)}\label{SVD}

The singular value decomposition  (SVD) \cite{golub1971singular} is one of the most important matrix decomposition algorithms in the past decades. The SVD provides a foundation for most of the data-driven methods that are currently being used, as it provides a numerically stable matrix decomposition that can be used for a variety of purposes.\\

Particularly, considering a data matrix $\bm{V}_1^K \in \mathbb{R}^{J\times K}$, such that:
\begin{equation} \label{Eq01}
\bm{ V}_1^K=[\bm{v}_1,\bm{v}_2,\dots , \bm{v}_K] , 
\end{equation}

where $\bm{v}_k$ is a reshaped snapshot (i.e. the \textit{k}th frame in any of the echocardiography) collected at time $t_k$, with $ k = 1,\dots , K $. Specifically, each column is a vector that contains the pixels $J = n_x \times n_y$ of the image analyzed. \\

The SVD algorithm permits the representation of the matrix $\bm{V}_1^K$ as a product of three other matrix factor as follows: 
\begin{equation}\label{Eq02}
\bm{V}_1^K = \bm{W} \bm{\Sigma}\bm{T}^T = \sum_{j=1}^{r = min(J,K)} \sigma_{j} \bm{w}_{j} \bm{t}_{j}^{T}, 
\end{equation}

where $\bm{W}$ $\in \bm{\mathbb{ C}}^{J\times J}$ and $\bm{T}$ $\in\bm{ \mathbb{C}}^{K\times K}$ are real, orthonormal matrices. The columns of $\bm{W}$ (noted $\bm{w}_j$) are the left singular vectors of $\bm{V}_1^K$, and the columns of $\bm{T}$ (noted $\bm{t}_j$) are the right singular vectors of $\bm{V}_1^K$. The matrix  $\bm{\Sigma} \in \mathbb{R}^{J\times K} $, which contains the singular values, is a matrix with real, non-negative entries on the diagonal (noted $\sigma_{j}$) and zeros off the diagonal. \\

Additionally, the SVD allows an optimal, low rank representation of the dataset using a rank of approximation $r'\leq r$. Hence, eq. (\ref{Eq02}) can be re-written as

\begin{equation}\label{Eq03}
\bm{V}_1^K \simeq \sum_{j=1}^{r'} \sigma_{j} \bm{w}_{j} \bm{t}_{j}^{T}. 
\end{equation}

This is an important property of the SVD, as high-dimensional data may be well described by a few dominant patterns given by the singular vectors.\\

The rank of approximation $r'$ can be chosen manually by specifying an integer value, or based on a given reconstruction error tolerance $\varepsilon_{SVD}$ as

\begin{equation}
\frac{\| \bm{V}^{K}_{1}-\sum_{j=1}^{r'} \sigma_{j} \bm{w}_{j} \bm{t}_{j}^{T}\|_F}{\sqrt{\sum_{j=1}^{r}\sigma_j^{2}}} \leq \varepsilon_{SVD},
\end{equation}

where $ \|.\|_F$ is the Frobenius norm.

\subsection{CNN architecture and training}
The convolutional neural network (CNN) chosen in this contribution have been designed for a categorical classification of images. This architecture of a CNN has proven to be one of the most commonly used and  robust image classification models (\cite{lee2017deep,acharya2018automated,sharma2018analysis,cai2020review,puttagunta2021medical}). In view of the success of this CNN architecture in image classification research, we decided to adapt this structure for our study. \\ 

In order to build the ideal model architecture for our dataset, hyperparameter tuning was performed. The tuning was done for 10 epochs with the max number of trials equals to 6, to chose number of convolutional layers, the units of the dense layers and the learning rate. The final architecture of the CNN is as follow: first, three ($3\times 3$) convolutional layers are included, where each one of these layers is followed by a ($ 2 \times 2$) max pooling layer. Next, a flatten layer is inserted and finally two dense fully connected layers are added. The two dense layers are accompanied with the activation functions Rectified Linear Unit (ReLU) and the soft-max, respectively. \\

    

In this contribution, the training was performed for 80 epochs, with a batch size of $128$, the chosen optimizer is RMSprop, the learning rate chosen based on the hyperparameter tuner was set to $10^{-3}$, and categorical-crossentropy as loss function. Finally, validation loss and validation accuracy are chosen to be performance metrics and the confusion matrix is used to determine the nature and the rate of misclassification.


\section{Material} \label{material}
\subsection{Echocardiography data}

\begin{figure*}[h!]

    \centering
\subfloat[ \label{Fig2_a}]{\includegraphics[width=8cm, height=7cm]{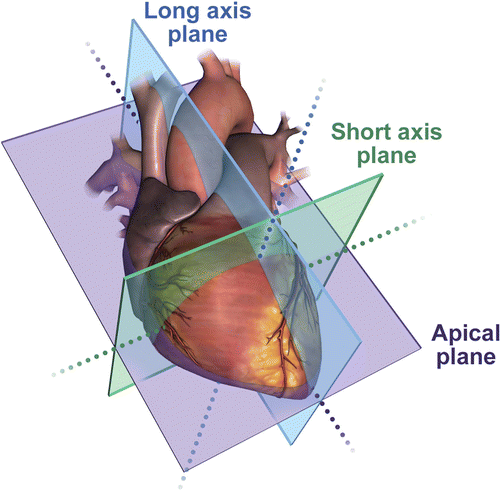}} \hspace{0.1cm} 
\subfloat[ \label{Fig2_b} ]{\includegraphics[width=8.5cm, height=7.5cm]{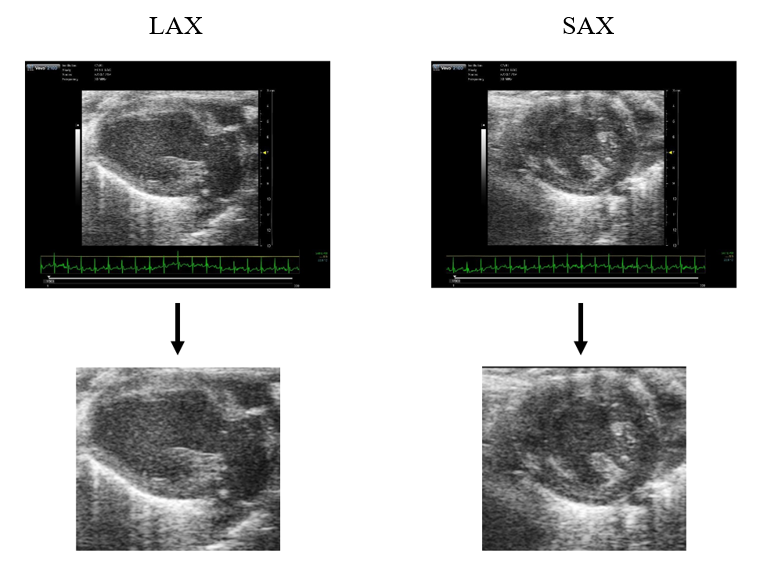}}  \\	
\caption{(a): Imaging planes of the heart: the long axis (LAX)  and the short axis (SAX) (original source of the image \cite{mitchell2019guidelines}). (b) Original echocardiography images before and after cropping.}
    \label{Fig2} 

\end{figure*}

This study involves 260 echocardiography datasets (videos), taken with respect to two distinct views: a long axis view (LAX, see long axis plane in Fig. (\ref{Fig2_a})) and short axis view (SAX, see short axis plane in Fig. (\ref{Fig2_a})). The 260 datasets split equally into 130 dataset for LAX and 130 datasets for SAX, where each dataset (video) consists of a number of frames varies between 90 to 120 frames (snapshots). In each case, the 130 datasets (which we will call samples) are equally distributed as 26 dataset per cardiac condition (which we will call class), where 20 samples are used for the training, and the remaining 6 samples are hold-out for evaluating the model on new, unseen data. \\

Figure (\ref{Fig2_b}) shows a sample of the original echocardiography images pre-cropping (top row)  and post-cropping (bottom row). Cropping is applied to all images depicting the various cardiac conditions, aimed at eliminating text, borders, ECG  information and emphasizing solely the heart in echocardiography images.\\    

For the classification process, the 100 training samples (20 samples per class), which are associated to the following cases: healthy (H), obesity (Ob), diabetic cardiomyopathy (DC), myocardial infarction (MI) and TAC hypertrophy (HT), are split as follow: 90 images of each sample are taken, giving a number of 1800 images per class. The resulting database, which consist of a total number of 9000 images, is divided into training, validation and testing/prediction (6000 images for training and 2500 for validation and 500 for prediction/testing). Similarly, 90 images from each one of 6 hold-out samples are taken, producing a  second testing database with a total number of 2700 images, which will be used to evaluate the performance of the model on \textit{new} (i.e. never seen by the classifier) data (table \ref{DATA} compiles the statistics for the datasets utilized in this study).\\

\renewcommand{\arraystretch}{1.5}
\begin{table}
\begin{center}
\begin{tabular}{|c|c|c|c||c|}
\cline{1-5}
 \multirow{2}{1.5cm}{Class}& \multicolumn{4}{c|}{\centering 
 Classification dataset } 

\bigstrut \\ \cline{2-5}
  & \multicolumn{1}{c|}{Training} & \multicolumn{1}{c|}{Validation} & \multicolumn{1}{c||}{Testing}  &\multicolumn{1}{c|}{Unseen data} 


\bigstrut \\ \cline{1-5} 
  H & 1200 & 500 & 100 & 540 \\ 
\cline{1-1} \cline{2-5}
  DC & 1200 & 500 & 100 & 540 \\ 
\cline{1-1} \cline{2-5}
  MI & 1200 & 500 & 100 & 540 \\ 
\cline{1-1} \cline{2-5}
  Ob & 1200 & 500 & 100 & 540 \\ 
\cline{1-1} \cline{2-5}
  HT & 1200 & 500 & 100 & 540 \\ 
\cline{1-1} \cline{2-5}

\end{tabular}
\caption{Number of images used for the classification for both LAX and SAX datasets. The acronyms used are  H: healthy, DC: diabetic cardiomyopathy, MI: myocardial infarction, Ob: obesity, HT: TAC hypertension.} \label{DATA}
\end{center}
\end{table}


    

\section{EigenHearts classification procedure}
In order to prepare the data for the classification process, a pre-processing stage was first carried out. This phase is accomplished in the  following steps:
\begin{itemize}
\item All the images from each cardiac condition are reshaped into vectors in $\mathbb{R}^J$, where $J = n_x \times n_y$, and arranged in an individual matrix  $\bm{V}_{C_i}$ with $i = 1, \dots , 5 $ (indexing the different health conditions), as shown in fig. (\ref{Fig3}).

\begin{figure}[h!]
    \centering
    \includegraphics[width=8.5cm, height=9cm]{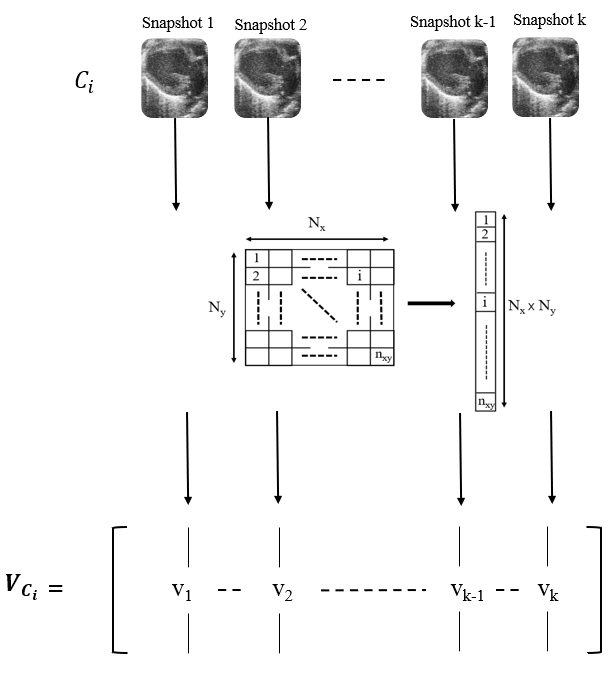}
    \caption{Schematic illustration of the process for preparing data matrices of the cardiac conditions.}
    \label{Fig3}
    
\end{figure}

\item The average heart of each cardiac condition is computed and subtracted from each column vector of its corresponding data matrix $\bm{V}_{C_i}$, giving $\tilde{\bm{V}}_{C_i}$.
\item The SVD is then applied separately to all the mean-subtracted matrices $\tilde{\bm{V}}_{C_i}$, which results in the following decomposition:

\begin{equation}
    \tilde{\bm{V}}_{C_i} = \bm{W_{C_i}}\bm{\Sigma}_{C_i}\bm{T}_{C_i}  \quad \textrm{with} \quad i = 1, \dots , 5 ,
    \end{equation}

where the columns of $\bm{W}_{C_i}$ represent ``the eigenhearts" related to each cardiac condition. Figure (\ref{Fig03}) shows the eigenhearts reshaped back into $ 256 \times 256 $ and illustrated as 2D images. 

    

\begin{figure*}[h!]

    \centering
\subfloat[ \label{Fig03_a}]{\includegraphics[width=9cm, height=10cm]{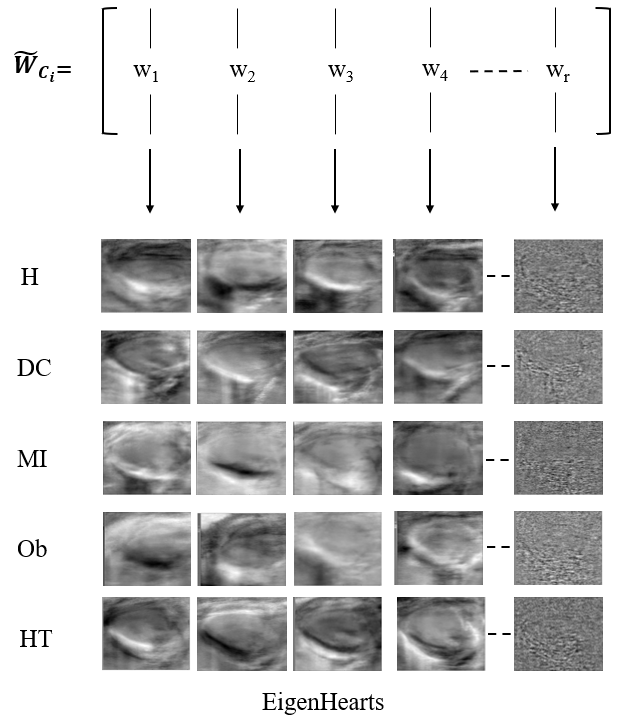}} \hspace{0.1cm} 
\subfloat[ \label{Fig03_b} ]{\includegraphics[width=9cm, height=10cm]{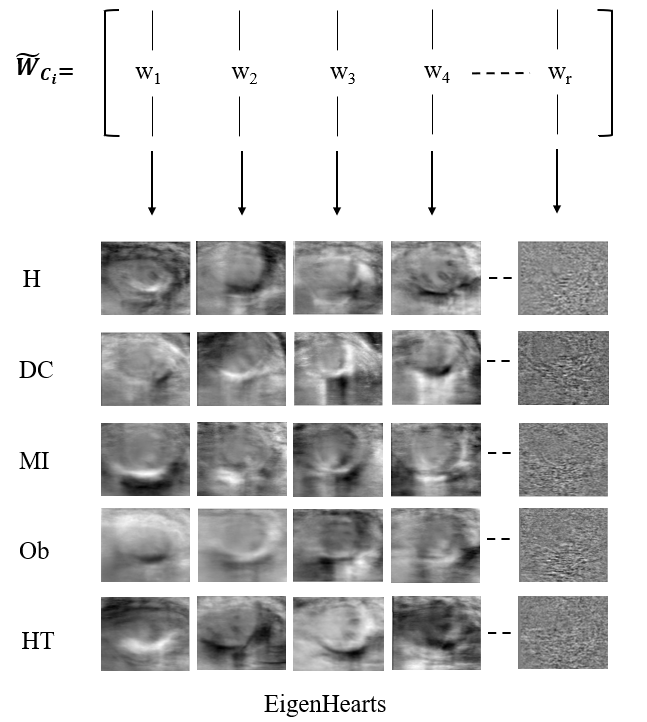}}  \\	
\caption{Eigenhearts of the five different cardiac conditions, (a) for the LAX datasets, (b) for SAX datasets. The acronyms used are  H: healthy, DC: diabetic cardiomyopathy, MI: myocardial infarction, Ob: obesity, HT: TAC hypertension.}
    \label{Fig03} 

\end{figure*}

\item Next, the matrices $\bm{W}_{C_i}$, which represent the libraries of eigenhearts, are used to approximately represent the original images, with different truncation ranks $r$. Hence, an original image $I$ (reshaped into vector form) is represented in the new coordinate system using the  projection operation:

\begin{equation} \label{Eq06}
 I_r =    \bm{\tilde{W}}_{C_i} \bm{\tilde{W}}^T_{C_i} I .
\end{equation}

\item All the images related to a specific cardiac conditions (which are reshaped into vectors) will be approximated by projecting them into their corresponding pod modes using the projection operation in  Eq. (\ref{Eq06}). This projection rereads the underlying characteristics of each dataset, as the eigenhearts can be seen as a set of features characterizing the variation between the echocardiography images of each cardiac condition. As a consequence, by using this projection,  each image associated to a certain condition will  be characterized by this set of features, providing a compact representation that strengthens the  connection between the images of each cardiac condition. 
\item After all the images representing the various cardiac conditions are projected into their corresponding pod modes, we reshape them back to 2D format and split them into training, validation, testing. The new, unseen datasets are also projected into this new coordinate system using the same libraries of  eigenhearts, where each group of images are projected into their corresponding pod modes. 
\item To ensure a fair comparison between the original echocardiography datasets and the newly generated databases (images projected into the  pod modes), we uphold consistent sample numbers across all classes and maintain the identical data distribution as introduced in Section \ref{material}.

\end{itemize}  

  

\subsection{Optimal truncation:}

\begin{figure*}[h!]

    \centering
\subfloat[ \label{Fig04_a} ]{\includegraphics[width=7.5cm, height=5.5cm]{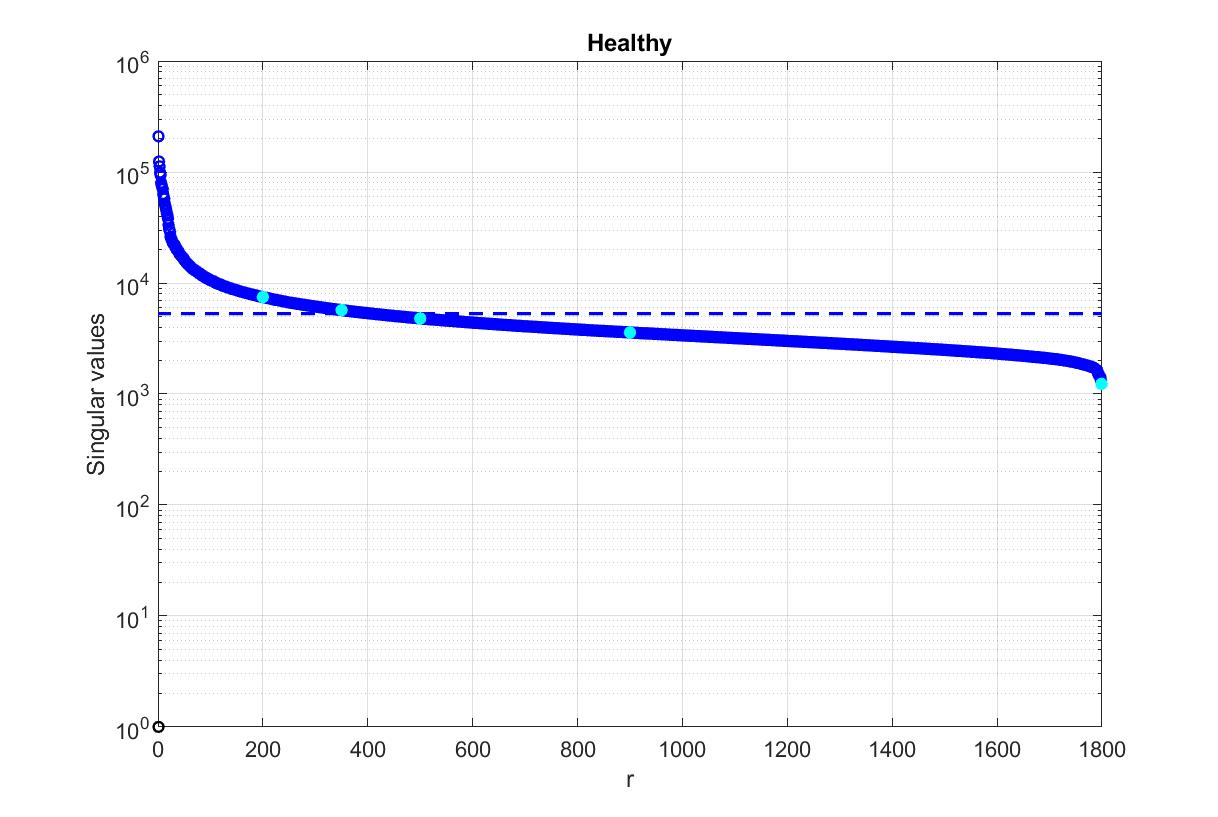}}  \\	 
\subfloat[ \label{Fig04_b}]{\includegraphics[width=7.5cm, height=5.5cm]{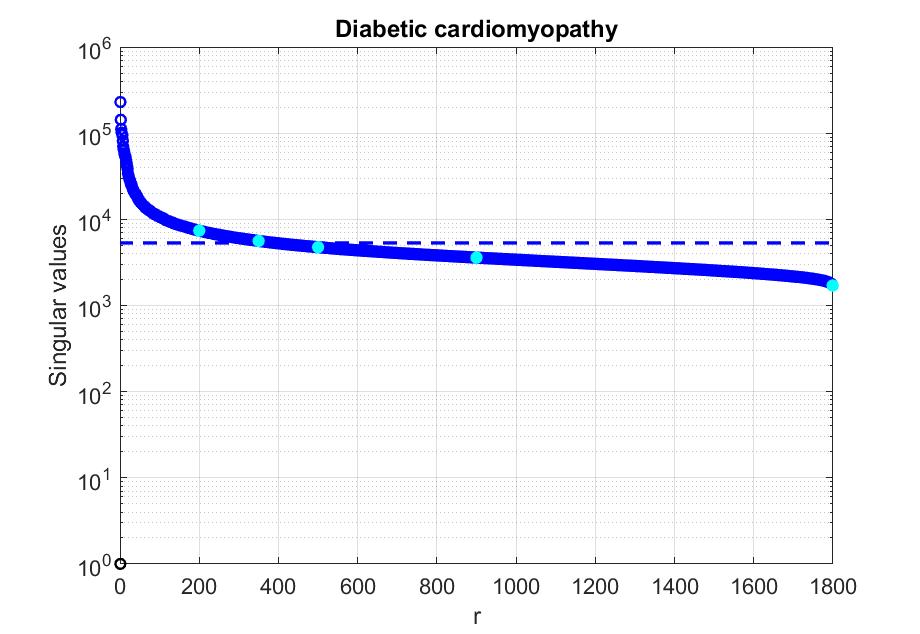}} \hspace{0.1cm} 
\subfloat[ \label{Fig04_c} ]{\includegraphics[width=7.5cm, height=5.5cm]{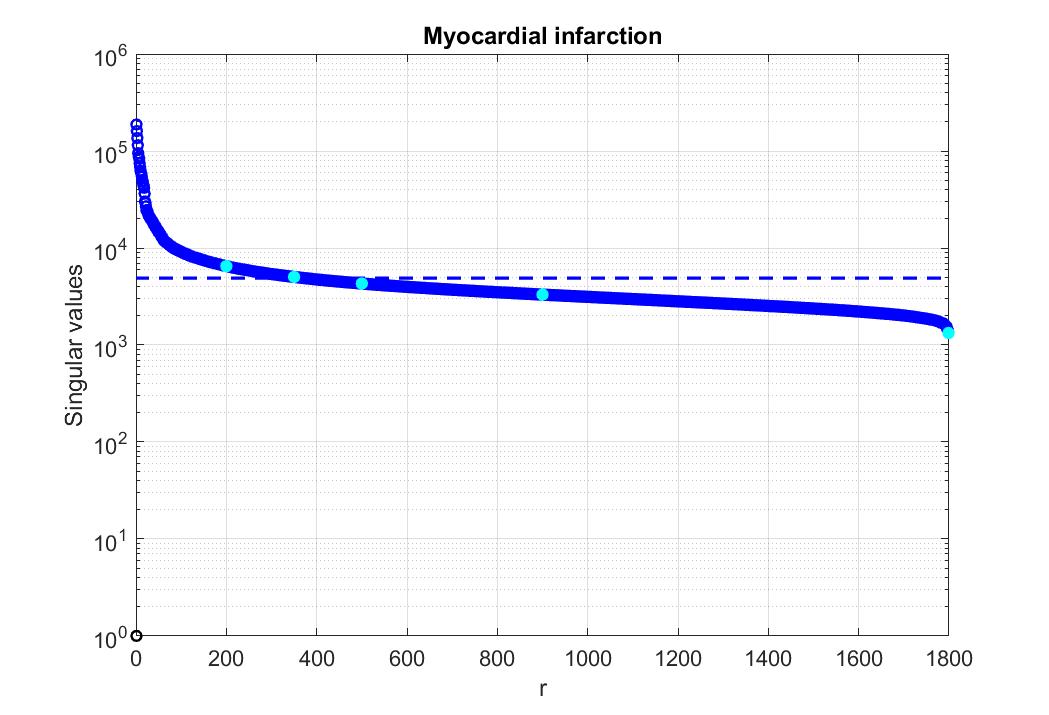}}  \\	
\subfloat[ \label{Fig04_d}]{\includegraphics[width=7.5cm, height=5.5cm]{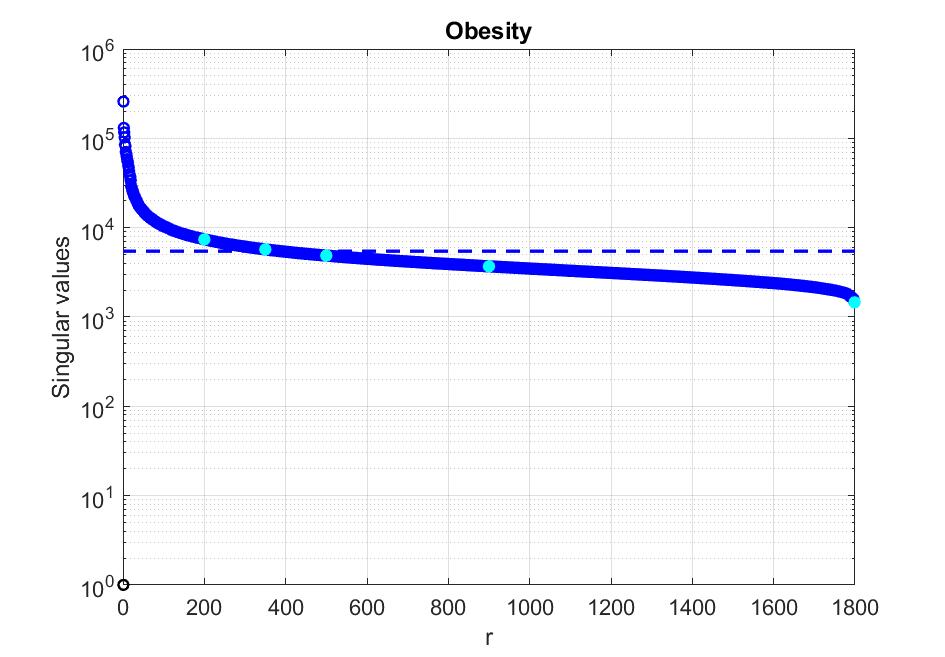}} \hspace{0.1cm} 
\subfloat[ \label{Fig04_e} ]{\includegraphics[width=7.5cm, height=5.5cm]{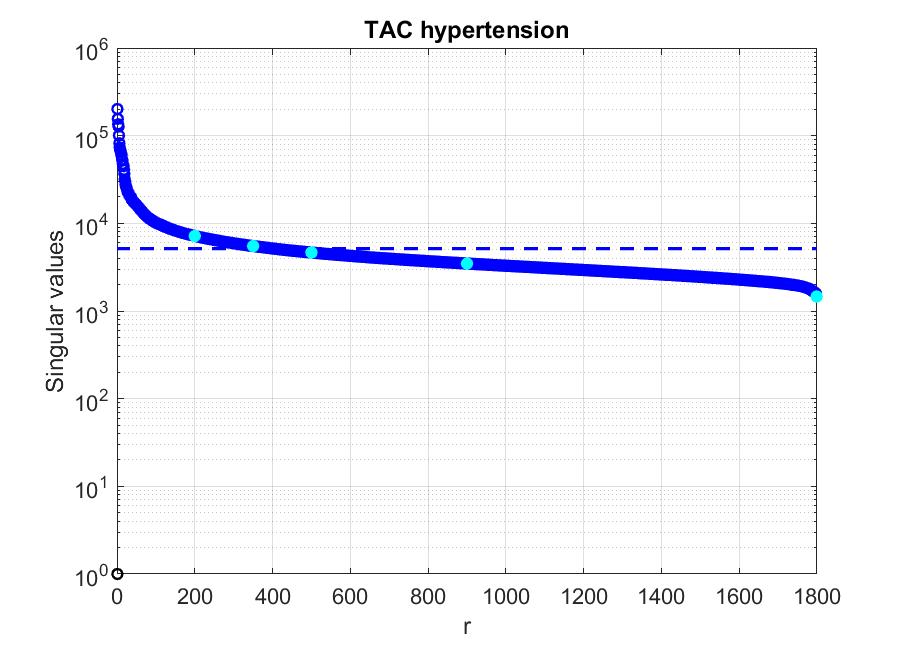}}  \\	

\caption{Distribution of the singular values for each data matrix for the LAX data. (a): healthy, (b): diabetic cardiomyopathy, (c): myocardial infarction, (d): obesity, (e): TAC hypertension. }
    \label{Fig04} 

\end{figure*}

\begin{figure*}[h!]

    \centering
\subfloat[ \label{Fig05_a} ]{\includegraphics[width=7.5cm, height=5.5cm]{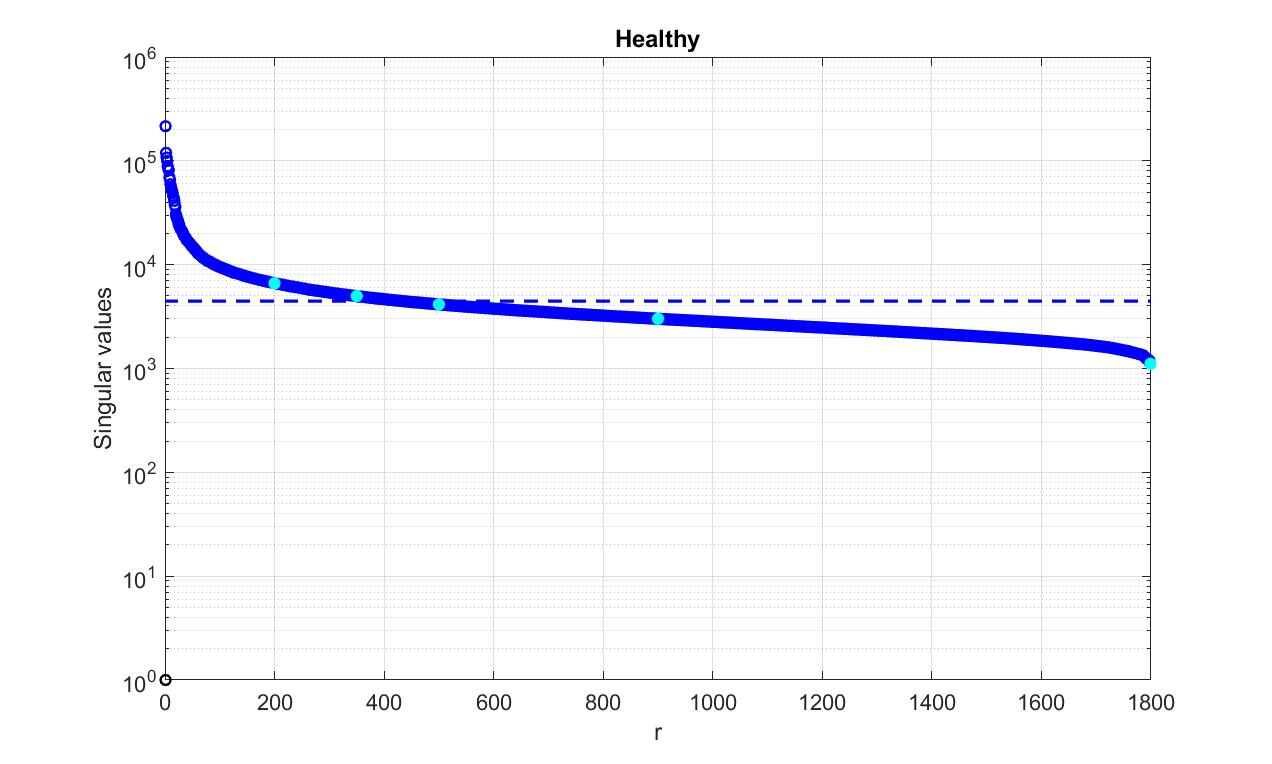}}  \\	 
\subfloat[ \label{Fig05_b}]{\includegraphics[width=7.5cm, height=5.5cm]{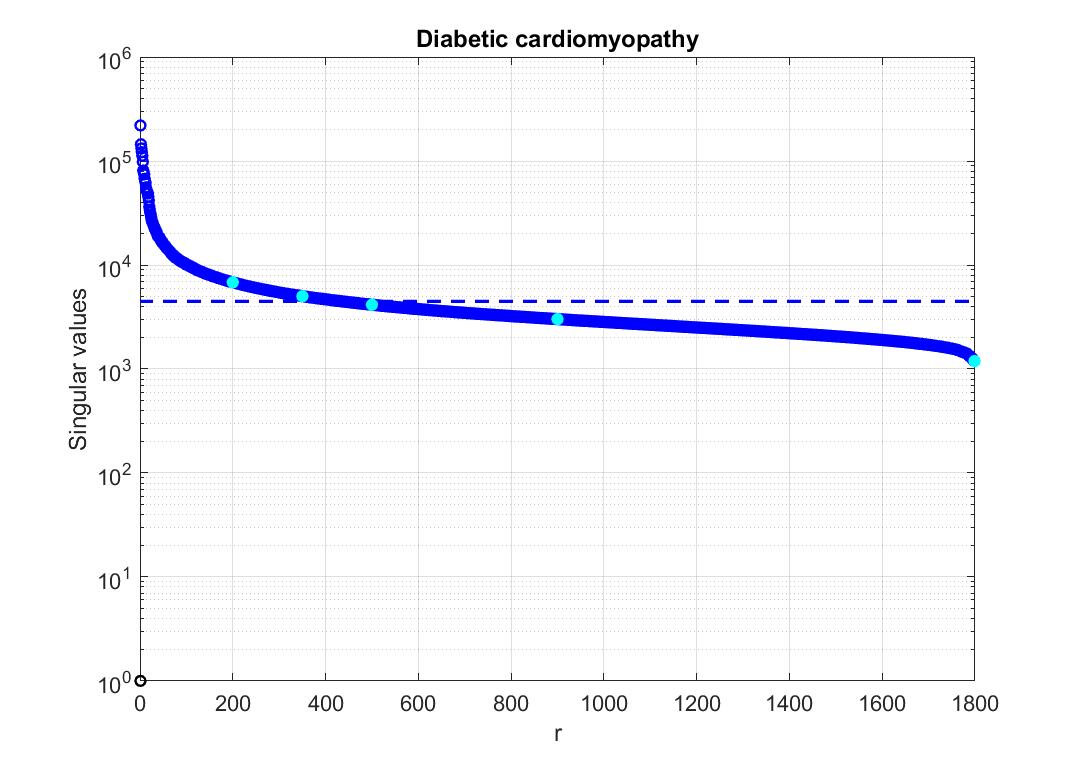}} \hspace{0.1cm} 
\subfloat[ \label{Fig05_c} ]{\includegraphics[width=7.5cm, height=5.5cm]{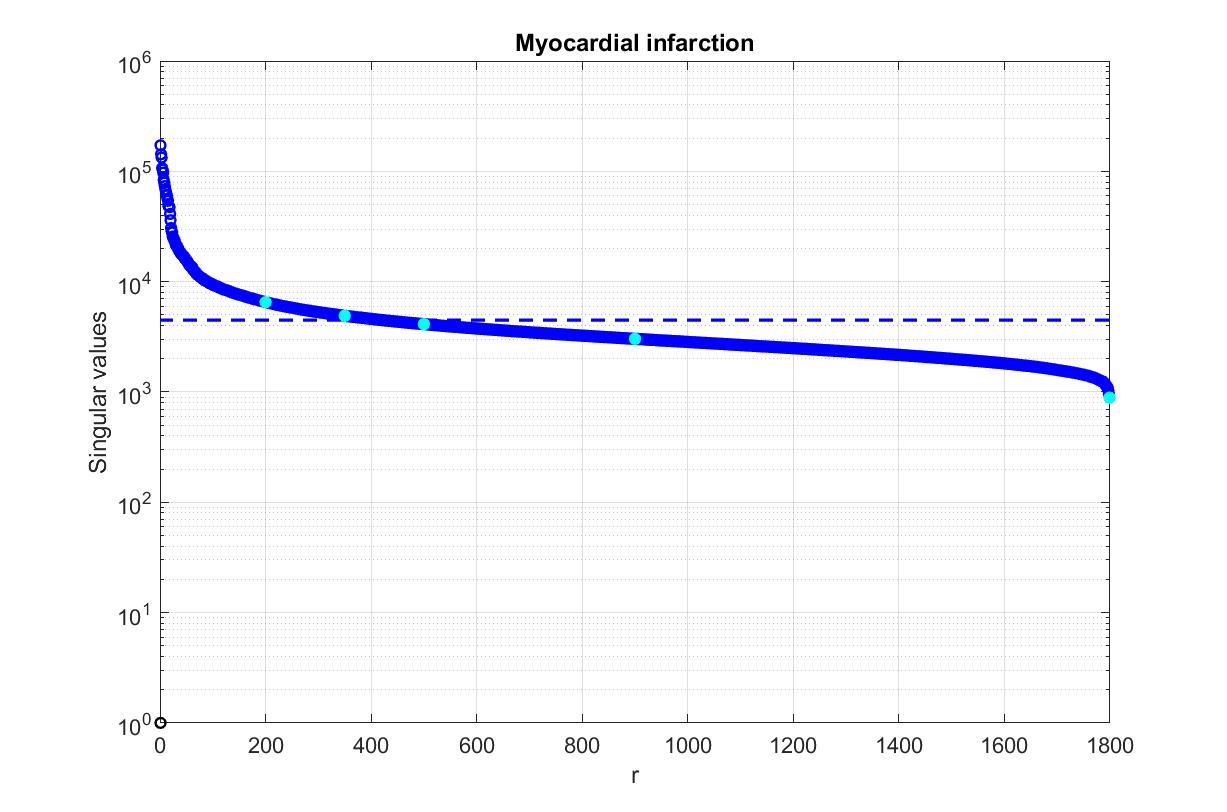}}  \\	
\subfloat[ \label{Fig05_d}]{\includegraphics[width=7.5cm, height=5.5cm]{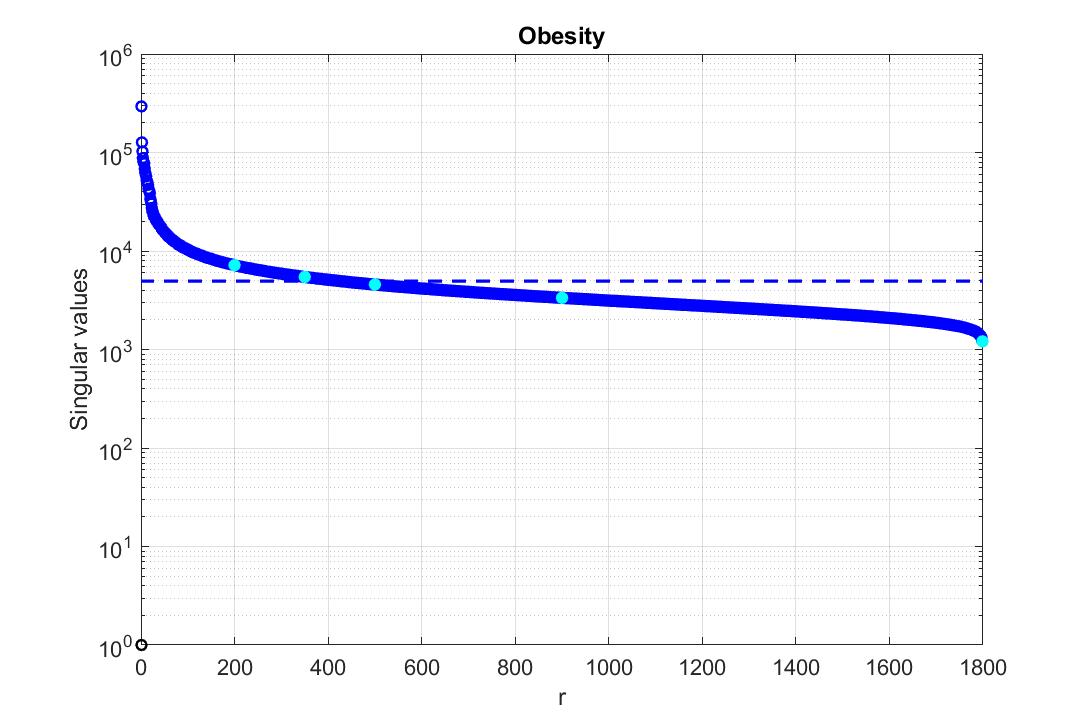}} \hspace{0.1cm} 
\subfloat[ \label{Fig05_e} ]{\includegraphics[width=7.5cm, height=5.5cm]{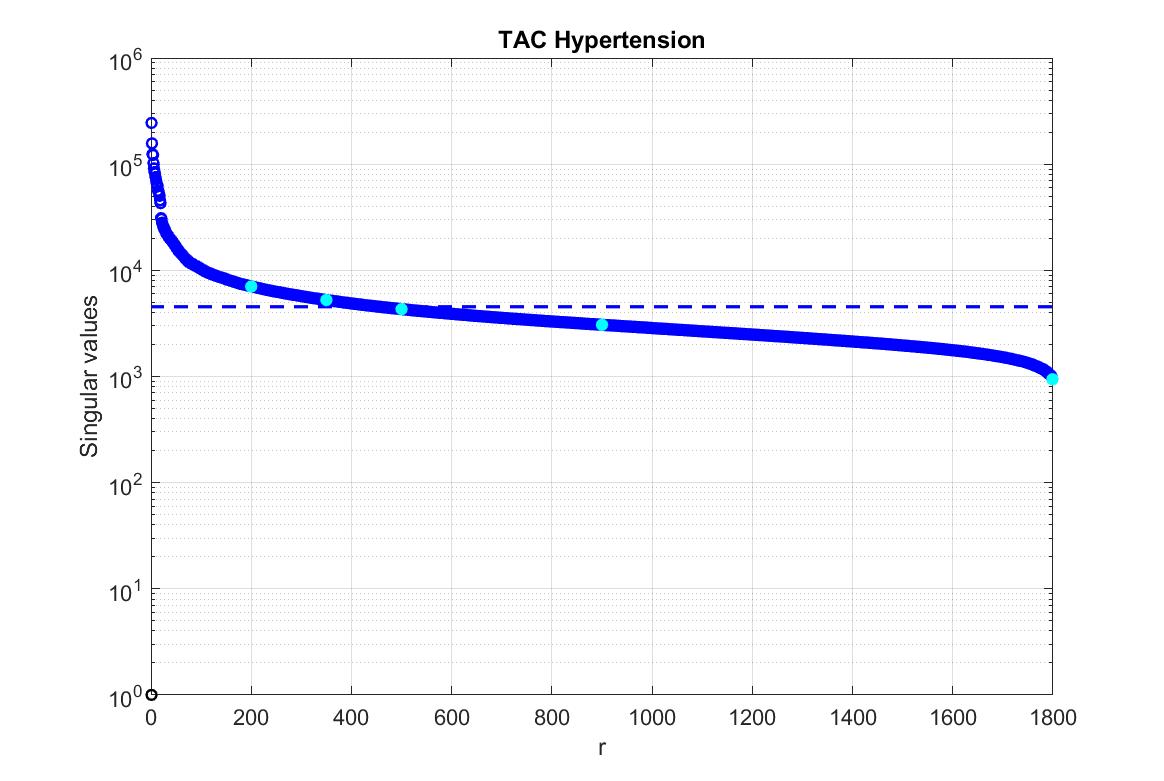}}  \\	

\caption{Distribution of the singular values for each data matrix for the SAX data. (a): healthy, (b): diabetic cardiomyopathy, (c): myocardial infarction, (d): obesity, (e): TAC hypertension. }
    \label{Fig05} 

\end{figure*}

Setting the optimal number of singular values to retain, known as truncation, stands as a critical and debated choice when employing the SVD algorithm. As this choice depends on several factors such as the noise level present in the data and singular value distributions. There are few approaches to select the truncation rank $r'$, including simply choosing a rank $r'$ that captures  a predefined proportion of variance or energy within the initial data, like $90\%$ or $99\%$ truncation. It is also common to identify any “elbows” or “knees” in the singular value distribution. This tactic is viewed as the shift from singular values denoting significant patterns to those indicative of noise. In this contribution, we alternatively explore the use of the Gavish and Donoho approach \cite{gavish2014optimal}. Their study extends from an extensive body of literature exploring different approaches to hard and soft thresholding of singular values. \\
In order to select the optimal threshold for our datasets, we apply the Gavish and Donoho to all the matrices $\bm{V}_{C_i}$ separately. Figures (\ref{Fig04}) and (\ref{Fig05}) shows the obtained distribution of the singular values of the different data matrices, where the optimal threshold is shown as the blue dashed line. Following these plots, we considered five truncation ranks $r'$, one prior to the threshold, set to $r' = 200$, one at the threshold $ r' = 350$, two after the threshold $r' = 500$  and $ r' = 900$ and finally a rank that captures $99\%$ of the information $r' = 1799$ . 

\section{Results} \label{Results}

To classify the different cardiac conditions, we first represent all the images (training, testing, validation and prediction) using the different ranks $r'$ and the projection formula  defined in  Eq. (\ref{Eq06}). Following the generation of these new databases, we proceed to train the CNN five times for each value of truncation $r$, where the results are compiled in tables  (\ref{LAX_Class_Sum}) and (\ref{SAX_Class_Sum}).
The metrics considered are:
\begin{itemize}
    \item The average accuracy: sum of all the obtained accuracy values divided by the number of values.
    \item  The standard deviation of the model's performance, which is calculated as
$$ stnd = \sqrt{\frac{\sum_{i=1}^{n}(Acc_i - mean)^2}{n-1}} , $$ 
where $Acc_i$ is the value of the  accuracy obtained in the \textit{i}th run, $mean$ is the average accuracy and $n$ is the total number of values.\\
Recall that a standard deviation close to zero indicates that accuracy values are very close to the mean accuracy, whereas a larger standard deviation indicates accuracy values are spread further away from the mean accuracy. 
\end{itemize}

To compare the performance of the proposed approach, the CNN is first trained using the original images (same data distribution explained in section \ref{material}). The results, presented in tables \ref{LAX_Class_Sum} and \ref{SAX_Class_Sum}, indicate modest performance when training the CNN with these original images. Despite achieving high accuracy during training (as reflected in validation accuracy), the performance on testing data is considered modest, with average accuracies of 81\% for LAX data and 76\% for SAX. Moreover, when predicting on new data, the accuracy levels are notably low, standing at 50\% for LAX and 43\% for SAX. \\
However, when repeating the experiment for the eigenhearts-enabled classification (i.e. employing the SVD for pre-processing), we observe significant and substantial improvements in classification accuracies. As depicted in tables \ref{LAX_Class_Sum} and \ref{SAX_Class_Sum} the validation accuracy still achieves high level at 100\% for both LAX and SAX data. However, there are distinct improvements in testing and prediction accuracies. As the optimal result for  LAX is attained for the truncation rank $r' = 200$, yielding a testing accuracy of 97\% and the prediction on the new data reached 98\%. Meanwhile, the best result for the SAX data is accomplished at $r' = 500$, with 93\% testing accuracy and 85\% for the prediction on new data.\\

These findings can be justifiable as follows: the LAX dataset exhibits reduced noise levels, allowing the SVD to effectively capture the most significant information in its initial modes. Consequently, utilizing just 200 modes proved adequate for image reconstruction, resulting in optimal classification accuracy. Notably, as the truncation rank increases (thus, allowing more noise to corupt the images), classification accuracy decreases.\\

In contrast, the SAX dataset contains comparatively higher noise levels. Furthermore, the images obtained from SAX view experience clear lack in discriminative information, thus necessitating a greater number of modes for accurate representation. As evidenced in table (\ref{SAX_Class_Sum}), fluctuations in accuracies are expected, meanwhile, the most favorable performance is observed at $r = 500$.  \\

\begin{table*}[h!]
\begin{center}
\begin{adjustbox}{max width=1\textwidth}
\begin{tabular}{l l l l l l l}
\toprule
\multirow{2}{3.5cm}{Accuracy}&  \multirow{2}{2.5cm}{Original images}& \multicolumn{5}{c}{\centering  Truncation}  \\
\cline{3-7} &  &\multicolumn{1}{l}{$r' = 200$} & \multicolumn{1}{l}{$r' = 350$} & \multicolumn{1}{l}{$r' = 500$} & \multicolumn{1}{l}{$r' = 900$} & \multicolumn{1}{l}{$r' = 1799$} \bigstrut \\ \hline
Validation  &  0.99 ± 0
    & $1 \pm 0 $ &    $1 \pm 0 $  & $1 \pm 0 $  & $1 \pm 0 $ & $1 \pm 0 $  \bigstrut \\ \hline
Testing  & 0.81 ± 0.058
 & 0.97 ± 0.027
 & 0.95 ± 0.029
 & 0.95 ± 0.012
 & 0.93 ± 0.033
 & 0.92 ± 0.031
  \bigstrut \\ \hline
Prediction (Unseen data)  &0.5 ± 0.041
 & 0.98 ± 0.021
 & 0.96 ± 0.025
 & 0.90 ± 0.017
 & 0.89 ± 0.051
 & 0.84 ± 0.35
  \bigstrut  \\ 
\bottomrule
\end{tabular} 
\end{adjustbox}
\caption{Accuracy results summary for the LAX data. The results are displayed as average accuracy ± standard deviation.} 
\label{LAX_Class_Sum}
\end{center}
\end{table*}

\begin{table*}[h!]
\begin{center}
\begin{adjustbox}{max width=1\textwidth}
\begin{tabular}{l l l l l l l}
\toprule
\multirow{2}{3.5cm}{Accuracy}&  \multirow{2}{2.5cm}{Original images}& \multicolumn{5}{c}{\centering  Truncation}  \\
\cline{3-7} &  &\multicolumn{1}{l}{$r' = 200$} & \multicolumn{1}{l}{$r' = 350$} & \multicolumn{1}{l}{$r' = 500$} & \multicolumn{1}{l}{$r' = 900$} & \multicolumn{1}{l}{$r' = 1799$} \bigstrut \\ \hline
Validation  &  $1 \pm 0 $    & $1 \pm 0 $ &    $1 \pm 0 $  & $1 \pm 0 $  & $1 \pm 0 $ & $1 \pm 0 $  \bigstrut \\ \hline
Testing  & 0.76 ± 0.13
 & 0.89 ± 0.036
 & 0.85 ± 0.066
 & 0.93 ± 0.016
 & 0.86 ± 0.11
 & 0.94 ± 0.030
  \bigstrut \\ \hline
Prediction (Unseen data)  & 0.43 ± 0.077
 & 0.82 ± 0.12
 & 0.78 ± 0.13
 & 0.85 ± 0.047
 & 0.77 ± 0.15
 & 0.81 ± 0.05
  \bigstrut  \\ 
\bottomrule
\end{tabular} 
\end{adjustbox}
\caption{Accuracy results summary for the SAX data. The results are displayed as average accuracy ± standard deviation.}
\label{SAX_Class_Sum}
\end{center}
\end{table*}

The results are detailed in the appendices (Fig\ref{Fig06} to Fig  \ref{Fig09}) through two methods. First, accuracy versus the number of epochs plots are utilized, showcasing both training and validation accuracies. This metric, deemed the most reliable, aligns with the evaluation of model performance, with accuracy being the preferred measure \cite{novakovic2017evaluation}. Second, the confusion matrix is employed to assess the model's performance when making predictions on either the testing set or new unseen data. This matrix displays model predictions against ground truth (true labels), illustrating correctly classified samples along the diagonal and misclassified samples off the diagonal. The results included, which can be seen in the appendices, are the ones concerning the training using the original images (in Fig \ref{Fig06} and Fig \ref{Fig08})and the ones associated with the truncation values that yielded the best performance ($ r' = 200$ for LAX presented in Fig. \ref{Fig07} and $ r' = 500$ for SAX presented in Fig \ref{Fig09}).

\FloatBarrier

\section{Conclusions} \label{concl}
In this contribution, we have explored the use of the famous EgienFaces approach for the classification of different cardiac conditions. In particular, the singular value decomposition (SVD) is used as a pre-processing step to reconstruct original echocardiography images. These modified representations of the echocardiography images are then utilized to train a convolutional neural network (CNN) for classifying and predicting five distinct cardiac conditions: healthy, diabetic cardiomyopathy, myocardial infarction, obesity, and TAC hypertension. The classification is achieved for two different databases: echocardiography images taken either  from the long axis view (LAX) or from the short axis view (SAX). To demonstrate  the efficiency of the proposed approach, we first train the CNN with the original echocardiography images. However, the performance of the classification model in this scenario is found to be modest, as the accuracy reached high levels only during the training with 99\% for LAX data and 100\% for SAX data. Nevertheless, the CNN's performance exhibits inconsistency,  as the accuracy dropped to 81\% for the LAX and 76\% for the SAX when evaluating the model on the testing data. Furthermore, the model experiences a major setback when predicting on unseen data, achieving only 50\% accuracy for LAX data and just 43\% for SAX data. On the contrary, when employing the SVD Eigenheart basis to reconstruct the echocardiography images, the model consistently demonstrates superior performance. As has shown the results, the accuracy during the training maintain the expected high levels at 100\% for both LAX and SAX datasets. Moreover, we observe a significant improvement in the testing and the prediction accuracies, as it increased to reach 97\% for LAX and 93\% for SAX for the testing and increased to 98\% for LAX and 85\% for SAX during the prediction on unseen data. Furthermore,  these outcomes are attained across various truncation values. As the significant performance was accomplished using truncation ranks $r' = 200$ and $r' = 500$ for LAX and SAX, respectively. This underscores the robust capabilities of the SVD algorithm in eliminating noise from the data and generating satisfactory images with minimal information, thereby significantly reducing computational costs.  \\

Drawing from the findings presented in this study, we can infer the effectiveness of the SVD algorithm and the EigenFaces approach in classifying cardiac diseases. This methodology notably improves the classification model's performance, resulting in an approximate 50\% increase in accuracy.
\section{Acknowledgments:}
The authors acknowledge the grants TED2021- 129774B-C21 and PLEC2022-009235 funded by MCIN/AEI/ 10.13039/501100011033 and by the European Union “NextGenerationEU”/PRTR and the grant PID2023-147790OB-I00 funded by MCIU/AEI/10.13039/501100011033/FEDER, UE. SLC acknowledges the MODELAIR and ENCODING projects that have received funding from the European Union’s Horizon Europe research and innovation program under the Marie Sklodowska-Curie grant agreement No. 101072559 and 101072779, respectively. The results of this publication reflect only the author(s) view and do not necessarily reflect those of the European Union. The European Union can not be held responsible for them.. Acknowledgments also to Agencia Española de investigation through  NextSim/AEI/10.13039/501100011033 and H2020, GA-956104.


\appendix
\section*{Appendices}\label{Appendices}


\clearpage
\newpage

\afterpage{%
\newgeometry{left=1cm,right=2cm,top=5mm, bottom=10mm}

\onecolumn




\begin{figure*}
\textbf{Appendix A. Accuracy plots and confusion matrices for classification using the original LAX data } \\
\vspace{0.2cm}

	\centering

\subfloat[  ]{\includegraphics[width=5cm, height=4.3cm]{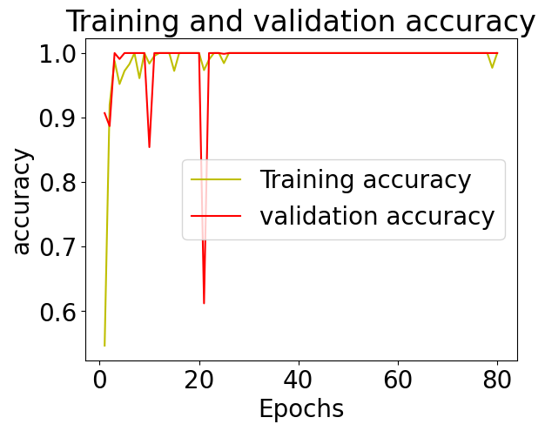}} \hspace{0.1cm} 
\subfloat[ ]{\includegraphics[width=5.8cm, height=4.3cm]{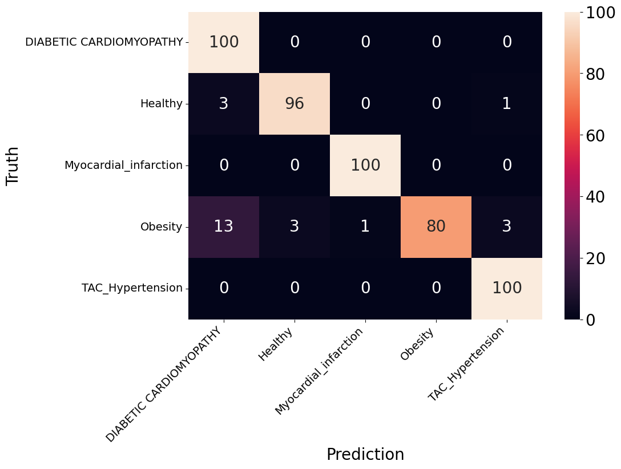}}\hspace{0.1cm} 
\subfloat[  ]{\includegraphics[width=5.8cm, height=4.3cm]{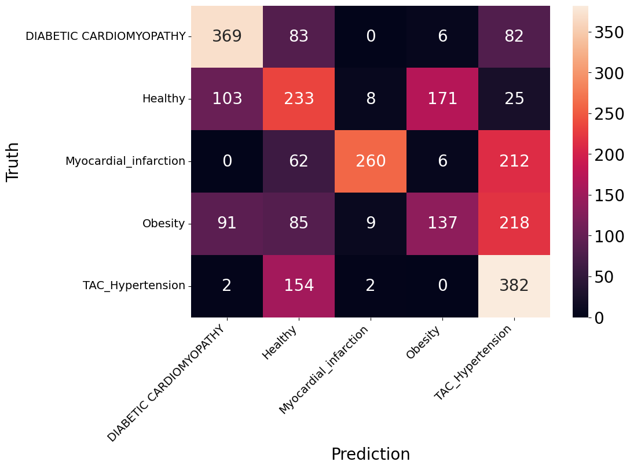}}\\
\subfloat[  ]{\includegraphics[width=5cm, height=4.3cm]{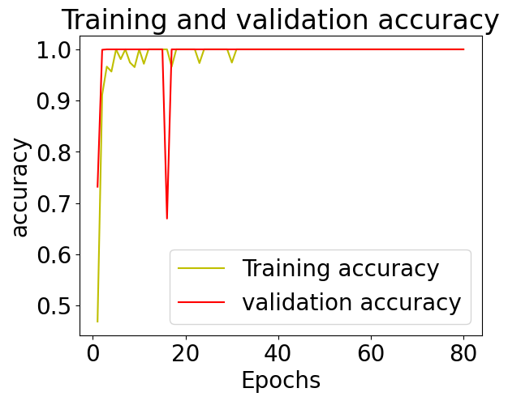}} \hspace{0.1cm} 
\subfloat[ ]{\includegraphics[width=5.8cm, height=4.3cm]{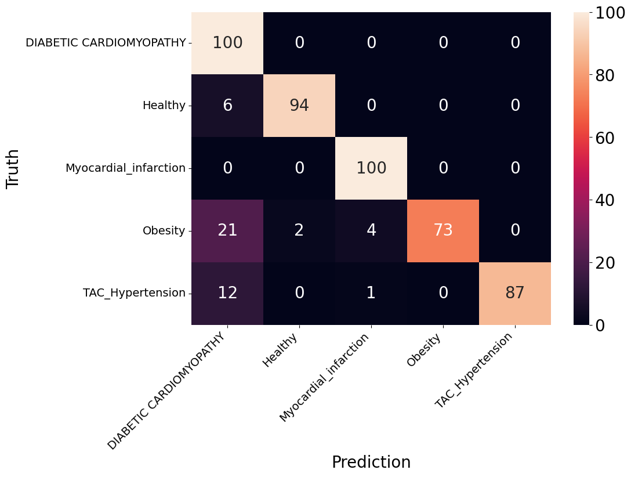}}\hspace{0.1cm} 
\subfloat[  ]{\includegraphics[width=5.8cm, height=4.3cm]{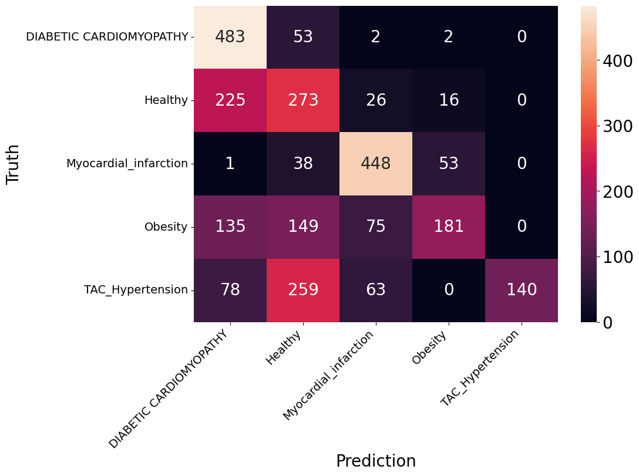}}\\

\subfloat[  ]{\includegraphics[width=5cm, height=4.3cm]{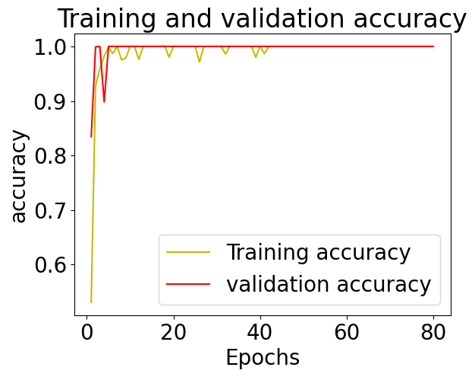}} \hspace{0.1cm} 
\subfloat[ ]{\includegraphics[width=5.8cm, height=4.3cm]{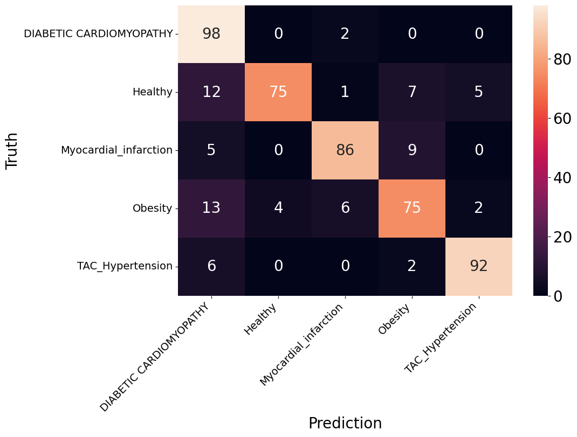}}\hspace{0.1cm} 
\subfloat[  ]{\includegraphics[width=5.8cm, height=4.3cm]{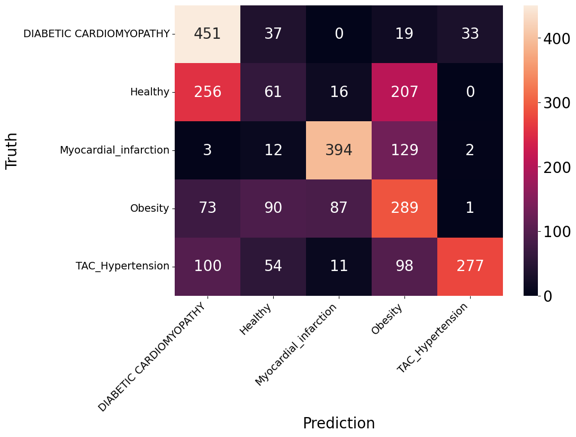}}\\

\subfloat[  ]{\includegraphics[width=5cm, height=4.3cm]{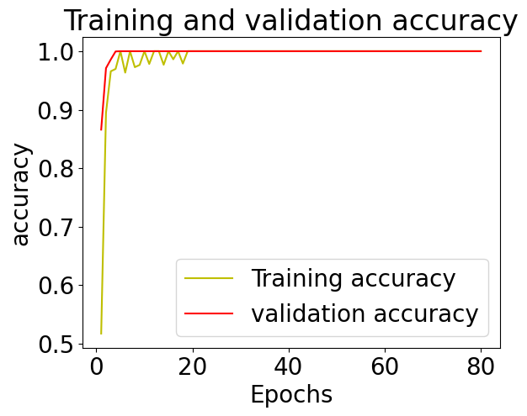}} \hspace{0.1cm} 
\subfloat[ ]{\includegraphics[width=5.8cm, height=4.3cm]{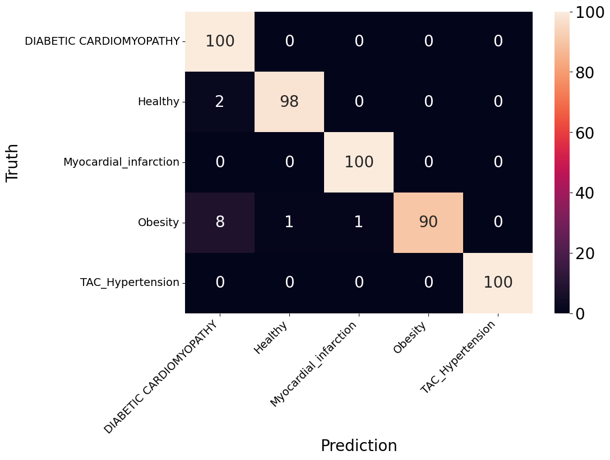}}\hspace{0.1cm} 
\subfloat[  ]{\includegraphics[width=5.8cm, height=4.3cm]{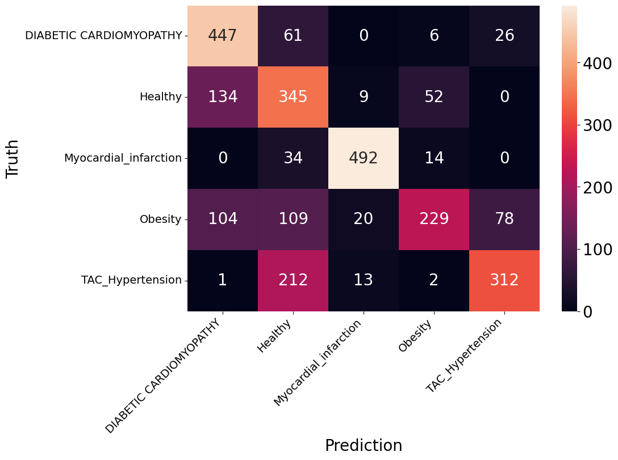}}\\

\subfloat[  ]{\includegraphics[width=5cm, height=4.3cm]{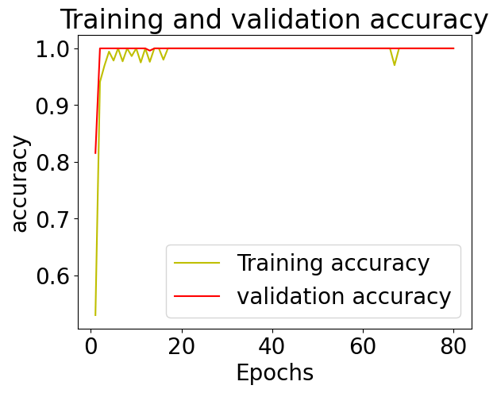}} \hspace{0.1cm} 
\subfloat[ ]{\includegraphics[width=5.8cm, height=4.3cm]{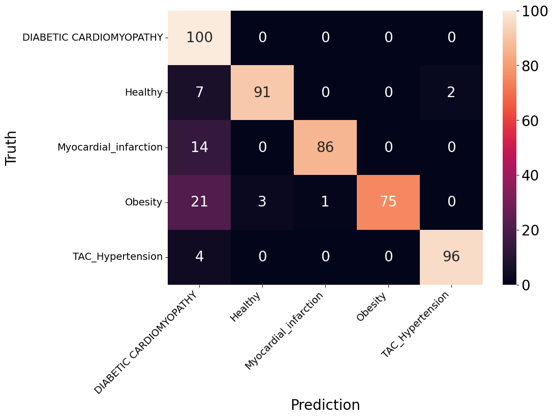}}\hspace{0.1cm} 
\subfloat[  ]{\includegraphics[width=5.8cm, height=4.3cm]{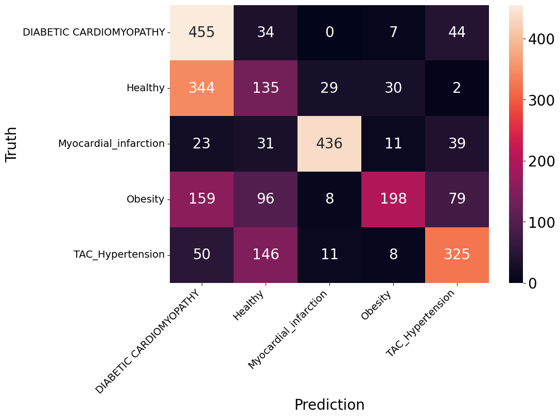}}\\

\centering
\captionsetup{format=plain,labelformat=simple}
\caption{Accuracy plot and confusion matrices illustrating the performance of the CNN when trained with the original images (LAX database). The first column displays accuracy during training, the second column presents confusion matrices of the testing phase, and the last column showcases confusion matrices of predictions on unseen data (recall that the model is trained 5 times).}	

	\label{Fig06}
    
\end{figure*}

\begin{figure}
\textbf{Appendix B. Accuracy plots and confusion matrices for the classification using LAX reconstructed data with $ r' = 200$ } \\
\vspace{0.2cm}

	\centering

\subfloat[  ]{\includegraphics[width=5cm, height=4.3cm]{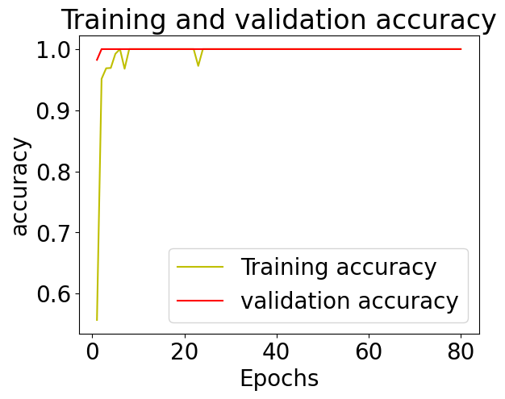}} \hspace{0.1cm} 
\subfloat[ ]{\includegraphics[width=5.8cm, height=4.3cm]{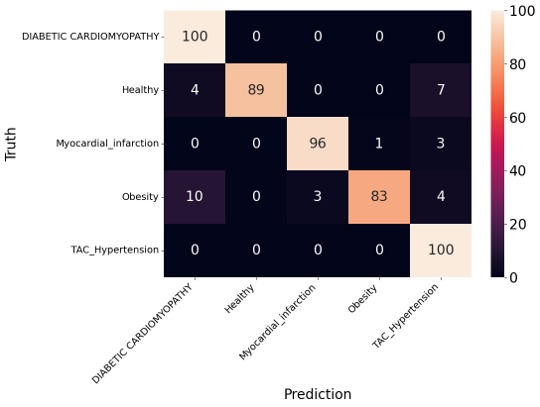}}\hspace{0.1cm} 
\subfloat[  ]{\includegraphics[width=5.8cm, height=4.3cm]{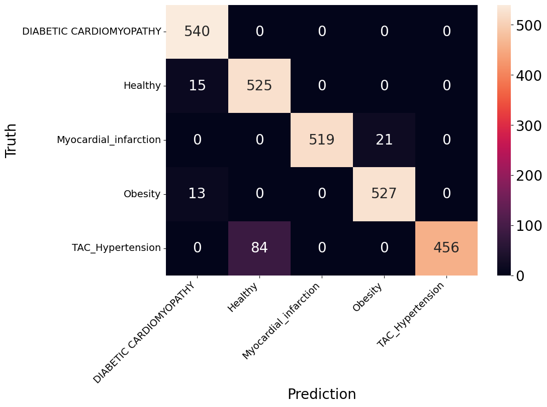}}\\
\subfloat[  ]{\includegraphics[width=5cm, height=4.3cm]{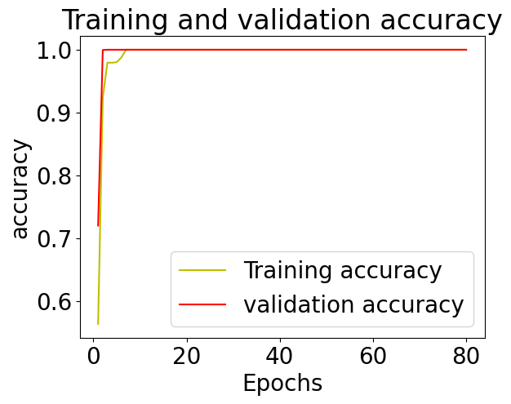}} \hspace{0.1cm} 
\subfloat[ ]{\includegraphics[width=5.8cm, height=4.3cm]{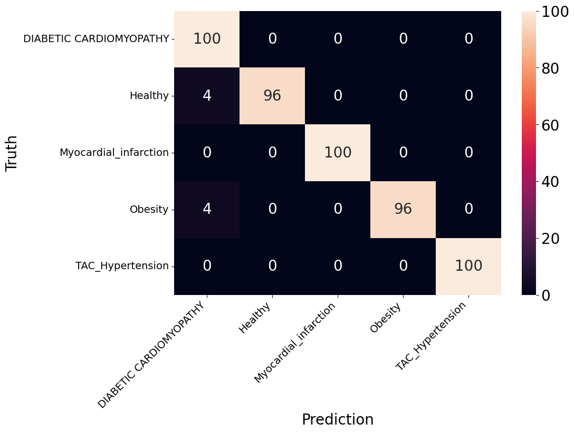}}\hspace{0.1cm} 
\subfloat[  ]{\includegraphics[width=5.8cm, height=4.3cm]{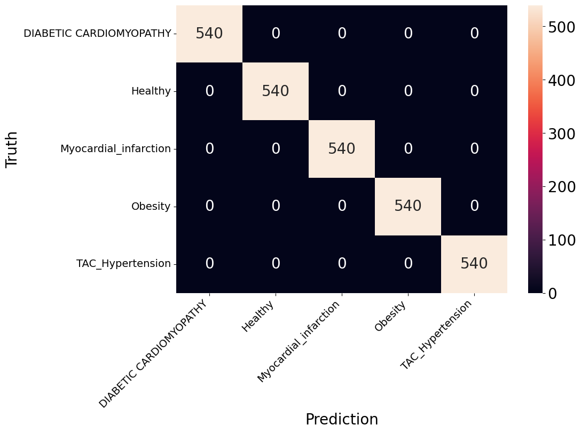}}\\

\subfloat[  ]{\includegraphics[width=5cm, height=4.3cm]{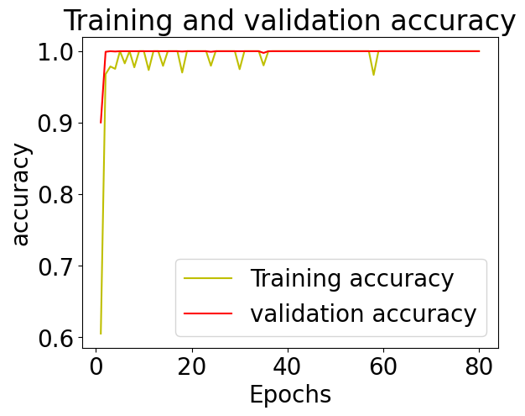}} \hspace{0.1cm} 
\subfloat[ ]{\includegraphics[width=5.8cm, height=4.3cm]{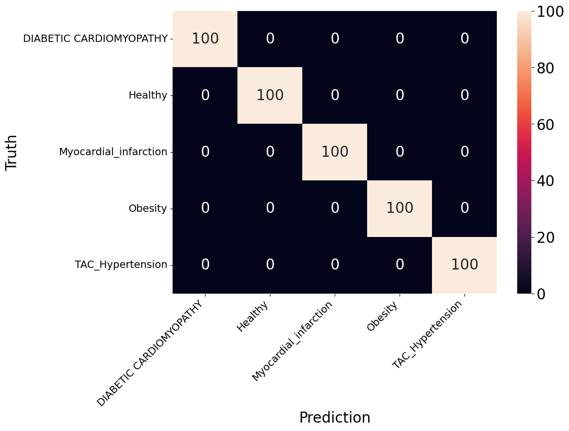}}\hspace{0.1cm} 
\subfloat[  ]{\includegraphics[width=5.8cm, height=4.3cm]{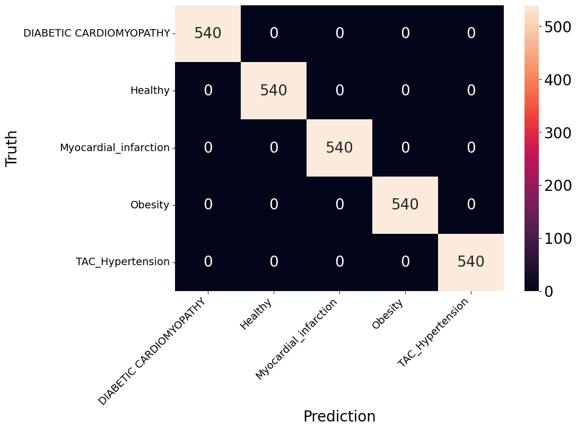}}\\

\subfloat[  ]{\includegraphics[width=5cm, height=4.3cm]{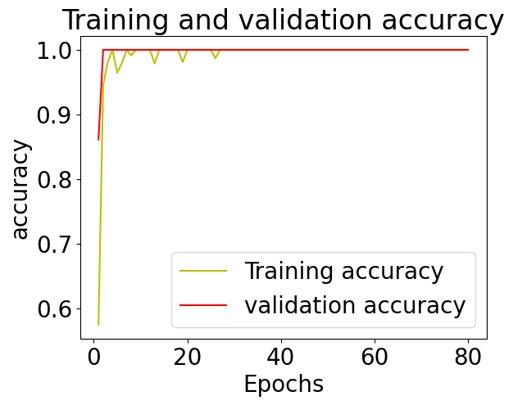}} \hspace{0.1cm} 
\subfloat[ ]{\includegraphics[width=5.8cm, height=4.3cm]{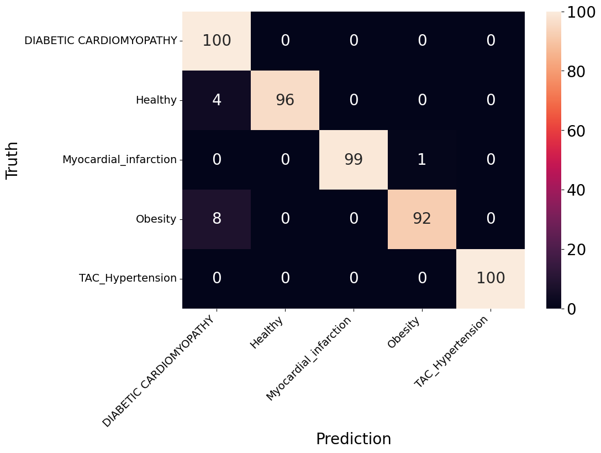}}\hspace{0.1cm} 
\subfloat[  ]{\includegraphics[width=5.8cm, height=4.3cm]{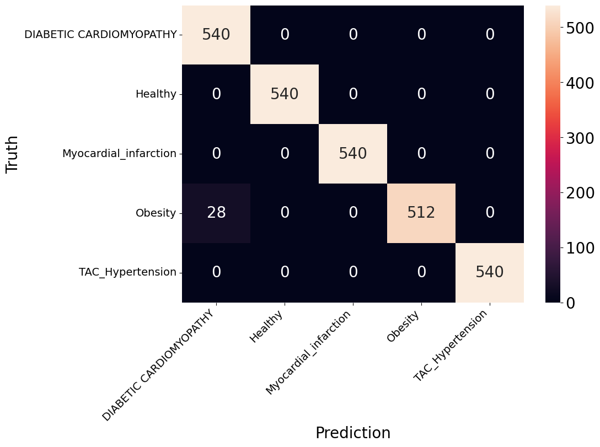}}\\

\subfloat[  ]{\includegraphics[width=5cm, height=4.3cm]{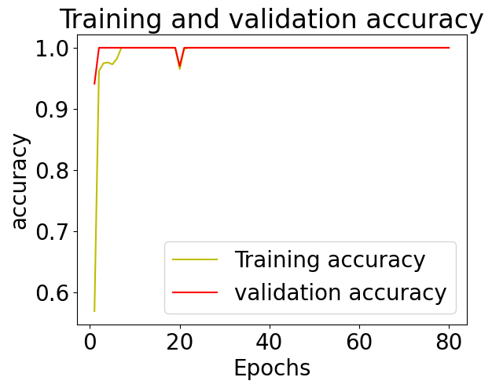}} \hspace{0.1cm} 
\subfloat[ ]{\includegraphics[width=5.8cm, height=4.3cm]{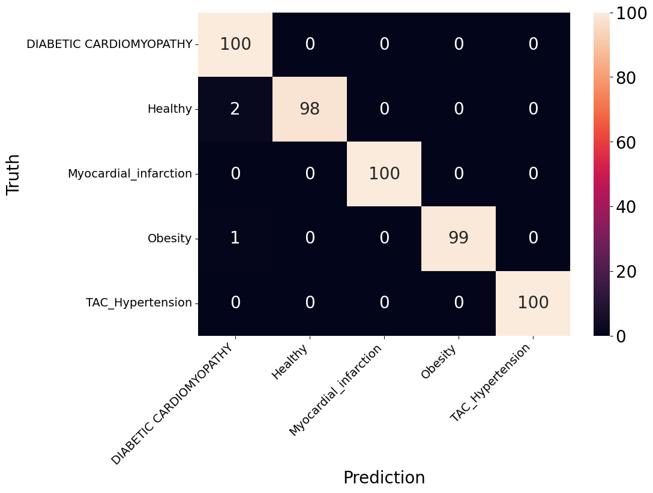}}\hspace{0.1cm} 
\subfloat[  ]{\includegraphics[width=5.8cm, height=4.3cm]{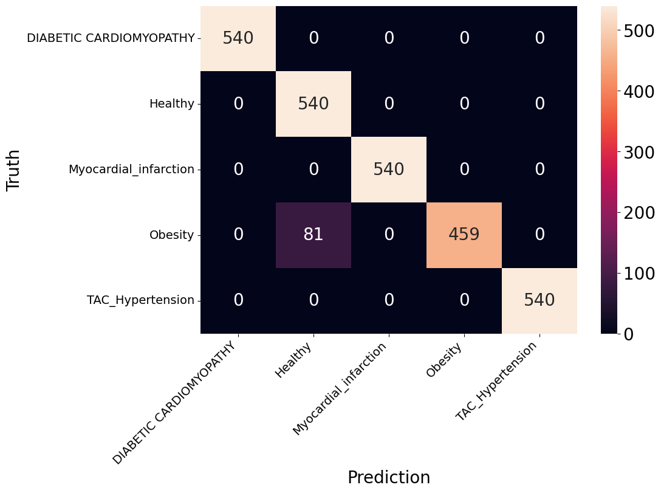}}\\

\centering
	\caption{Accuracy plot and confusion matrices illustrating the performance of the CNN when trained with the SVD reconstructed images with $r' = 200$ (LAX database). The first column displays accuracy during training, the second column presents confusion matrices of the testing phase, and the last column showcases confusion matrices of predictions on unseen data (recall that the model is trained 5 times).}	
	\label{Fig07}
\end{figure}

\begin{figure}
\textbf{Appendix C. Accuracy plots and confusion matrices for the classification using the original SAX data } \\
\vspace{0.2cm}

	\centering

\subfloat[  ]{\includegraphics[width=5cm, height=4.3cm]{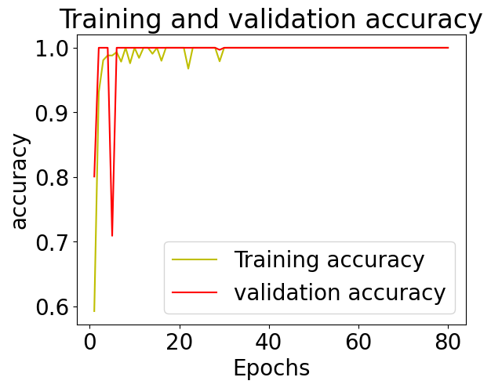}} \hspace{0.1cm} 
\subfloat[ ]{\includegraphics[width=5.8cm, height=4.3cm]{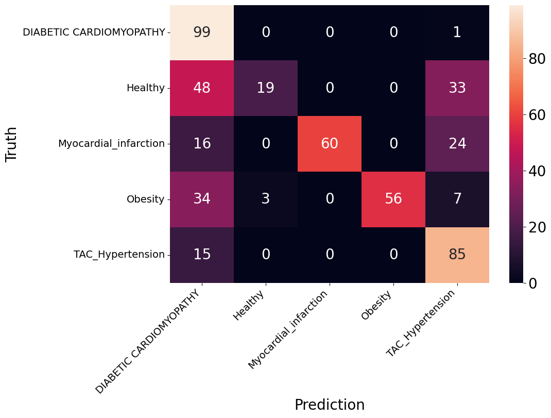}}\hspace{0.1cm} 
\subfloat[  ]{\includegraphics[width=5.8cm, height=4.3cm]{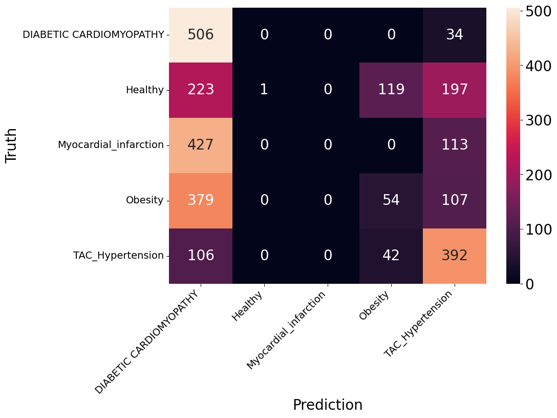}}\\
\subfloat[  ]{\includegraphics[width=5cm, height=4.3cm]{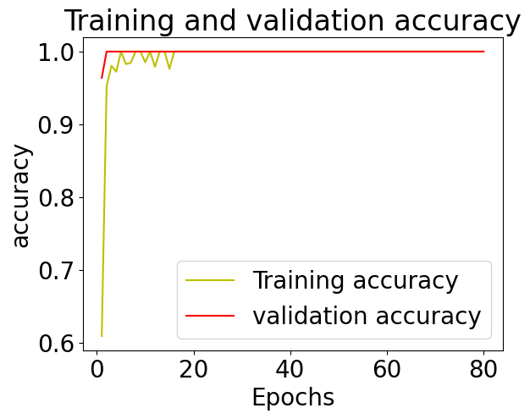}} \hspace{0.1cm} 
\subfloat[ ]{\includegraphics[width=5.8cm, height=4.3cm]{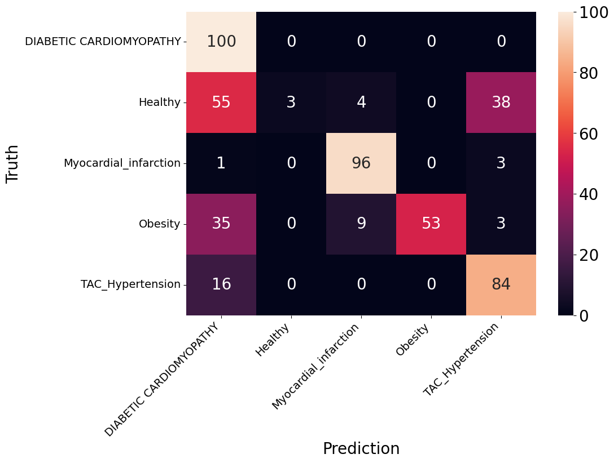}}\hspace{0.1cm} 
\subfloat[  ]{\includegraphics[width=5.8cm, height=4.3cm]{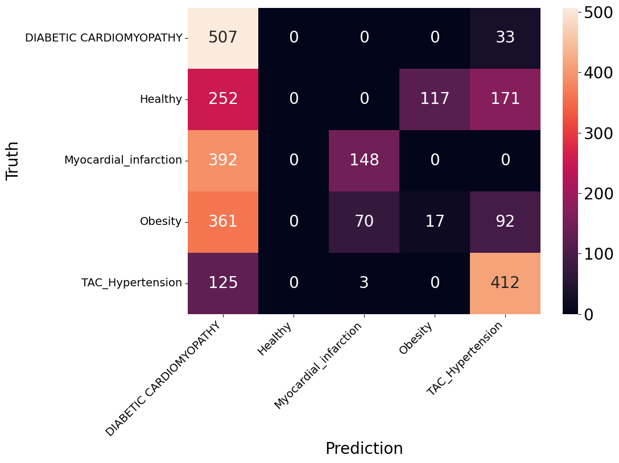}}\\

\subfloat[  ]{\includegraphics[width=5cm, height=4.3cm]{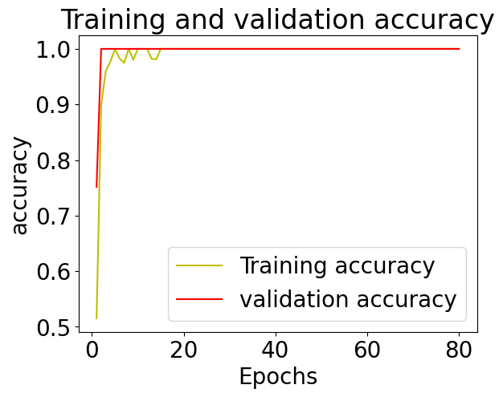}} \hspace{0.1cm} 
\subfloat[ ]{\includegraphics[width=5.8cm, height=4.3cm]{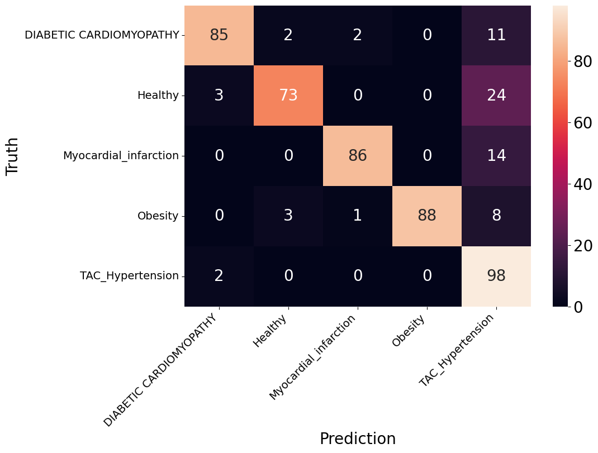}}\hspace{0.1cm} 
\subfloat[  ]{\includegraphics[width=5.8cm, height=4.3cm]{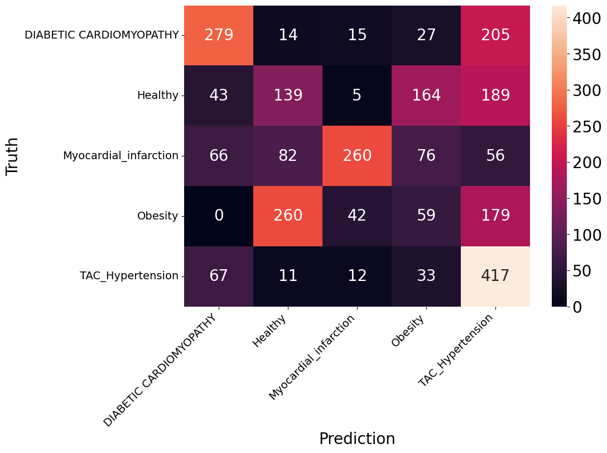}}\\

\subfloat[  ]{\includegraphics[width=5cm, height=4.3cm]{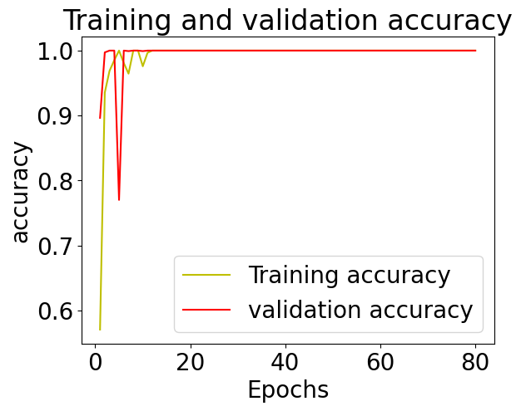}} \hspace{0.1cm} 
\subfloat[ ]{\includegraphics[width=5.8cm, height=4.3cm]{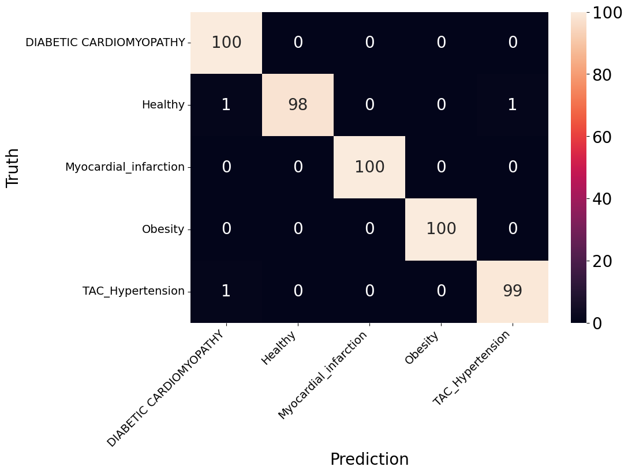}}\hspace{0.1cm} 
\subfloat[  ]{\includegraphics[width=5.8cm, height=4.3cm]{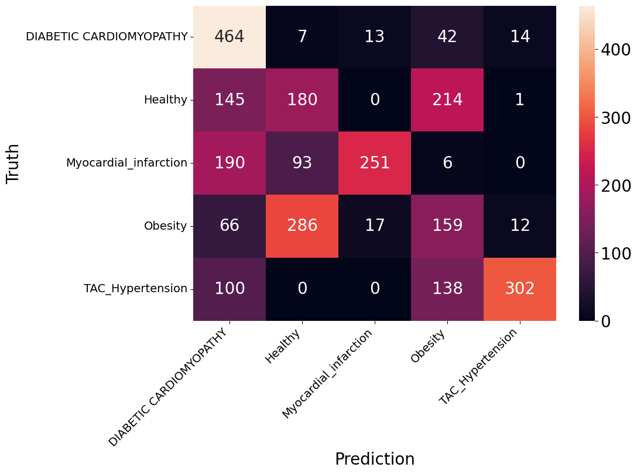}}\\

\subfloat[  ]{\includegraphics[width=5cm, height=4.3cm]{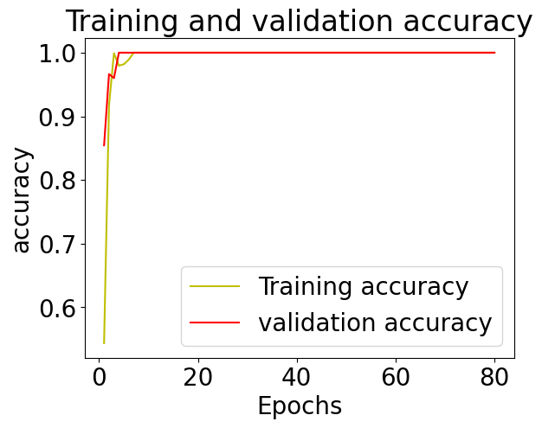}} \hspace{0.1cm} 
\subfloat[ ]{\includegraphics[width=5.8cm, height=4.3cm]{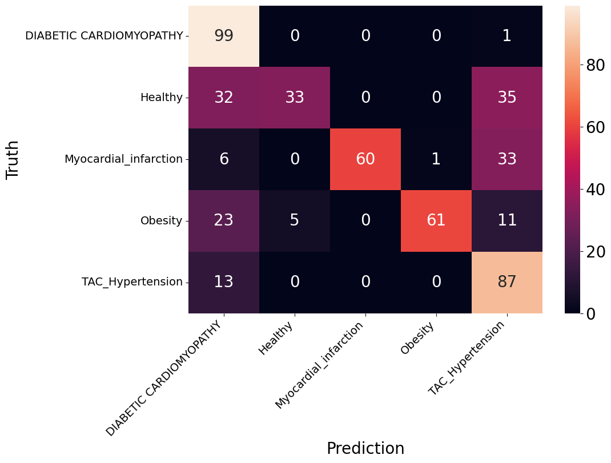}}\hspace{0.1cm} 
\subfloat[  ]{\includegraphics[width=5.8cm, height=4.3cm]{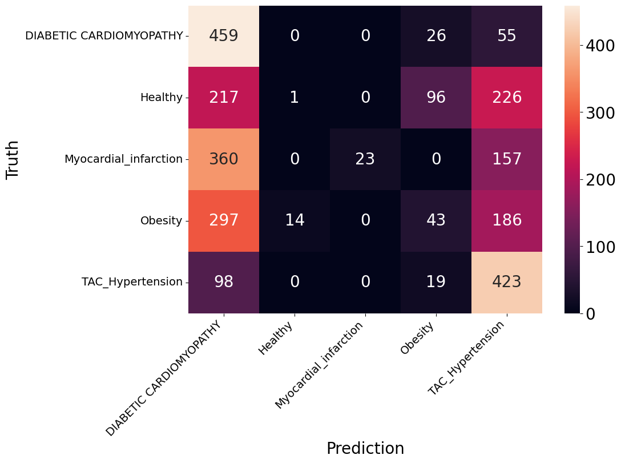}}\\

\centering
	\caption{Accuracy plot and confusion matrices illustrating the performance of the CNN when trained with the original images (SAX database). The first column displays accuracy during training, the second column presents confusion matrices of the testing phase, and the last column showcases confusion matrices of predictions on unseen data (recall that the model is trained 5 times).   }	
	\label{Fig08}
\end{figure}

\begin{figure}
\textbf{Appendix D. Accuracy plots and confusion matrices for the classification using SAX reconstructed data with $ r' = 500$ } \\
\vspace{0.2cm}

	\centering

\subfloat[  ]{\includegraphics[width=5cm, height=4.3cm]{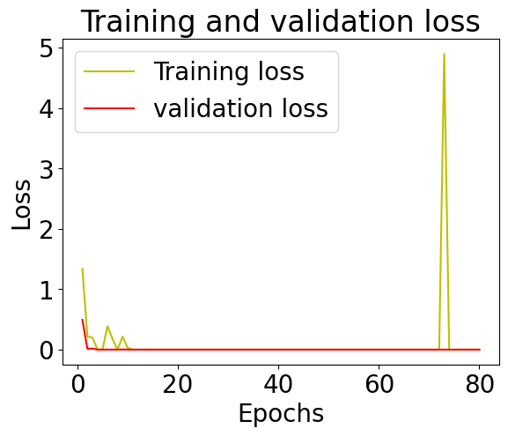}} \hspace{0.1cm} 
\subfloat[ ]{\includegraphics[width=5.8cm, height=4.3cm]{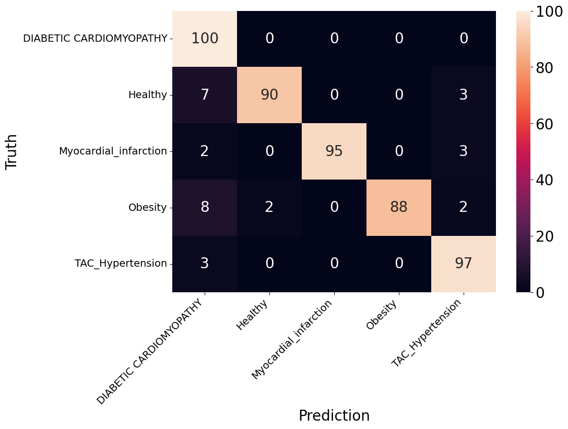}}\hspace{0.1cm} 
\subfloat[  ]{\includegraphics[width=5.8cm, height=4.3cm]{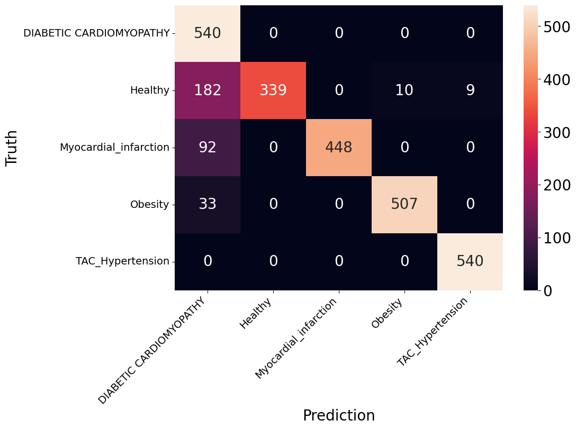}}\\
\subfloat[  ]{\includegraphics[width=5cm, height=4.3cm]{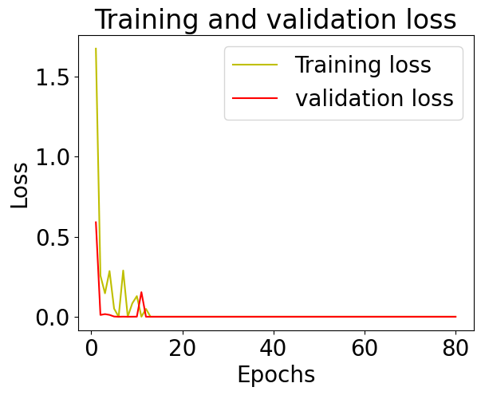}} \hspace{0.1cm} 
\subfloat[ ]{\includegraphics[width=5.8cm, height=4.3cm]{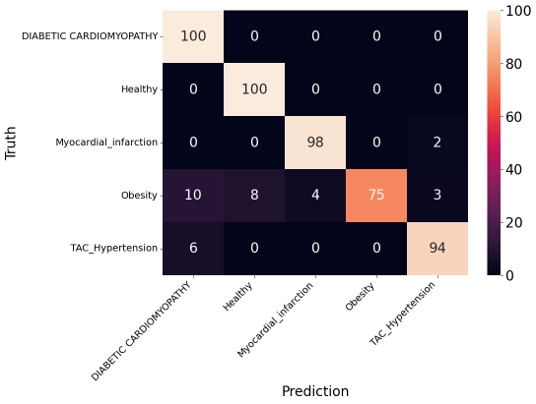}}\hspace{0.1cm} 
\subfloat[  ]{\includegraphics[width=5.8cm, height=4.3cm]{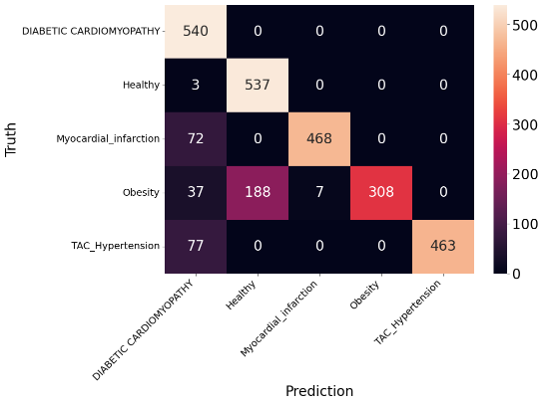}}\\

\subfloat[  ]{\includegraphics[width=5cm, height=4.3cm]{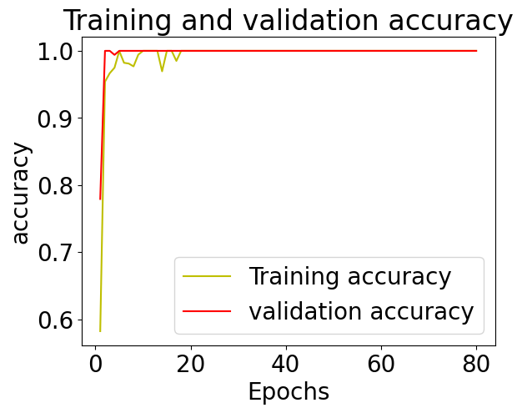}} \hspace{0.1cm} 
\subfloat[ ]{\includegraphics[width=5.8cm, height=4.3cm]{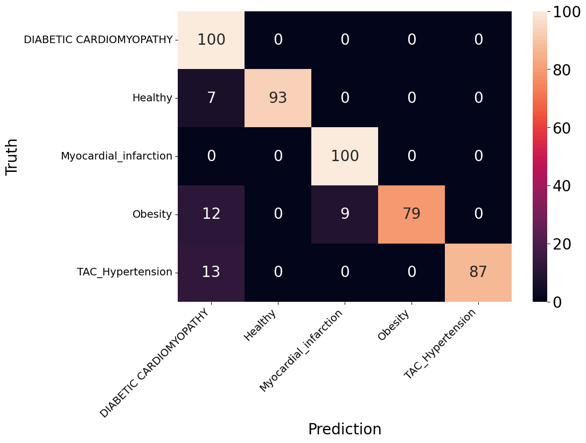}}\hspace{0.1cm} 
\subfloat[  ]{\includegraphics[width=5.8cm, height=4.3cm]{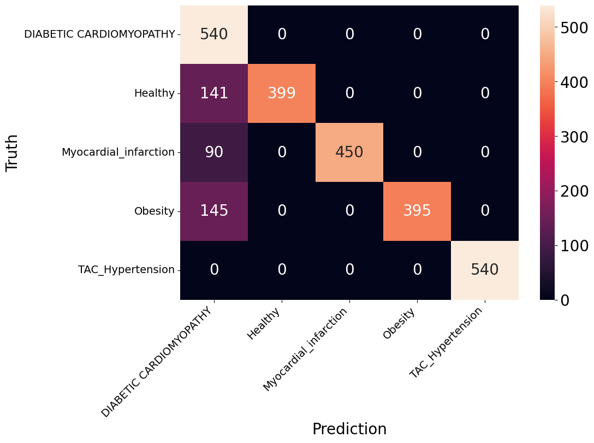}}\\

\subfloat[  ]{\includegraphics[width=5cm, height=4.3cm]{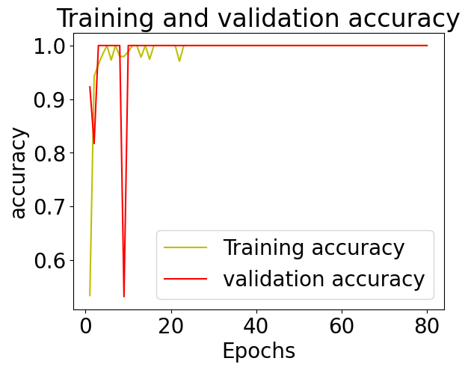}} \hspace{0.1cm} 
\subfloat[ ]{\includegraphics[width=5.8cm, height=4.3cm]{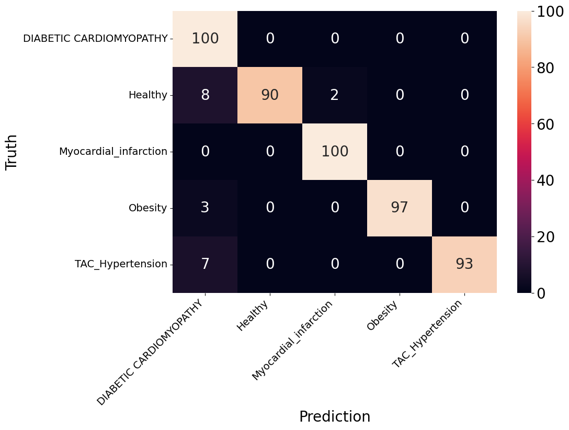}}\hspace{0.1cm} 
\subfloat[  ]{\includegraphics[width=5.8cm, height=4.3cm]{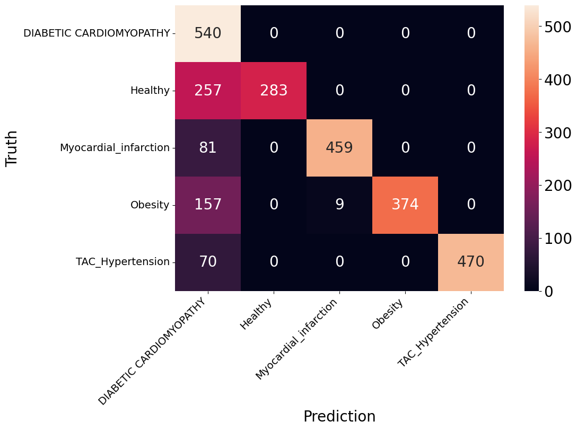}}\\

\subfloat[  ]{\includegraphics[width=5cm, height=4.3cm]{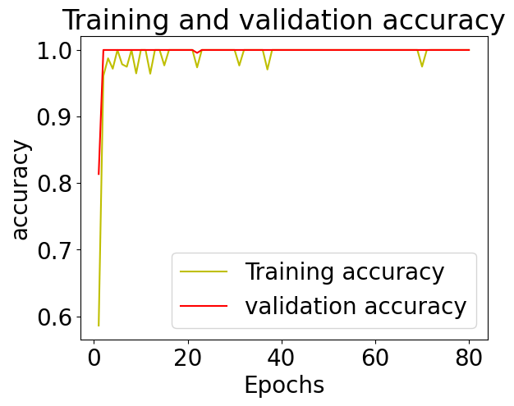}} \hspace{0.1cm} 
\subfloat[ ]{\includegraphics[width=5.8cm, height=4.3cm]{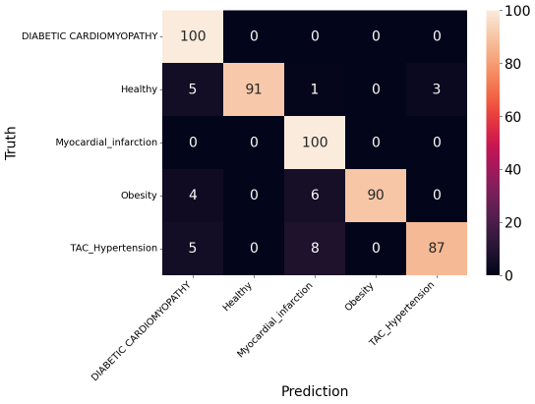}}\hspace{0.1cm} 
\subfloat[  ]{\includegraphics[width=5.8cm, height=4.3cm]{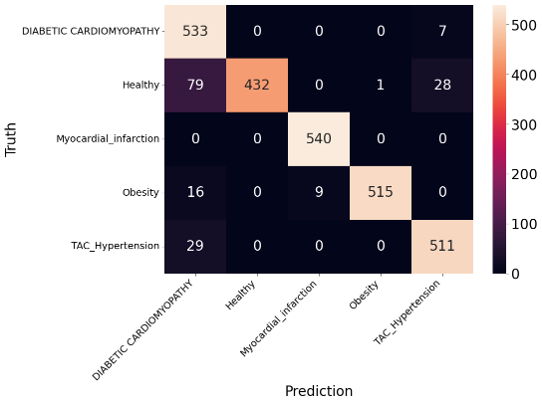}}\\

\centering
	\caption{Accuracy plot and confusion matrices illustrating the performance of the CNN when trained with the SVD reconstructed images with $r' = 500$ (SAX database). The first column displays accuracy during training, the second column presents confusion matrices of the testing phase, and the last column showcases confusion matrices of predictions on unseen data (recall that the model is trained 5 times).   }	
	\label{Fig09}
\end{figure}

\clearpage
\restoregeometry
}

\clearpage
\newpage

\afterpage{%
\newgeometry{left=1cm,right=2cm,top=2cm, bottom=2cm}



\twocolumn
 \bibliographystyle{elsarticle-num} 
 \bibliography{My_Refs}

\clearpage
\restoregeometry
}





\end{document}